\documentclass[aps,prd,nofootinbib,onecolumn,a4paper]{revtex4}
\usepackage[colorlinks,linkcolor=blue,anchorcolor=violet,citecolor=red]{hyperref}
\usepackage{amsmath,amssymb}
\usepackage[dvipdfmx]{graphicx}
\usepackage{float}
\usepackage{comment}
\usepackage{xcolor}
\usepackage{bm,latexsym,amsmath,amssymb,amsfonts,mathrsfs}
\usepackage{bbold}
\usepackage{ascmac}
\usepackage{physics}
\usepackage{color,float}
\usepackage{enumerate}

\definecolor{darkpastelgreen}{rgb}{0.01, 0.75, 0.24}
\newcommand{\cm}[1]{\textcolor{red}{#1}}

\begin{document}
\title{\large Quantumness of gravitational field: A perspective on monogamy relation}

\author{Yuuki Sugiyama}
 \email{sugiyama.yuki@phys.kyushu-u.ac.jp}
\affiliation{Department of Physics, Kyushu University, 744 Motooka, Nishi-Ku, Fukuoka 819-0395, Japan}

\author{Akira Matsumura}
 \email{matsumura.akira@phys.kyushu-u.ac.jp}
\affiliation{Department of Physics, Kyushu University, 744 Motooka, Nishi-Ku, Fukuoka 819-0395, Japan}
 
\author{Kazuhiro Yamamoto}
 \email{yamamoto@phys.kyushu-u.ac.jp}
\affiliation{Department of Physics, Kyushu University, 744 Motooka, Nishi-Ku, Fukuoka 819-0395, Japan}
\affiliation{
Research Center for Advanced Particle Physics, Kyushu University, 744 Motooka, Nishi-ku, Fukuoka 819-0395, Japan}
\affiliation{
International Center for Quantum-Field Measurement Systems for Studies of the Universe and Particles (QUP), KEK, Oho 1-1, Tsukuba, Ibaraki 305-0801, Japan}

\begin{abstract}
Understanding the phenomenon of quantum superposition of gravitational fields induced by massive quantum particles is an important starting point for quantum gravity.
The purpose of this study is to deepen our understanding of the phenomenon of quantum superposition of gravitational fields.
To this end, we consider a trade-off relation of entanglement (monogamy relation) in a tripartite system consisting of two massive particles and a gravitational field that may be entangled with each other. 
Consequently, if two particles cannot exchange information mutually, they are in a separable state, and the particle and gravitational field are always entangled.
Furthermore, even when two particles can send information to each other, there is a trade-off between the two particles and the gravitational field.
We also investigate the behavior of the quantum superposition of the gravitational field using quantum discord.
We find that quantum discord increases depending on the length scale of the particle superposition.
Our results may help understand the relationship between the quantization of the gravitational field and the meaning of the quantum superposition of the gravitational field.
\end{abstract}
\maketitle

\section{Introduction}
The unification of the gravity theory and quantum mechanics (quantum gravity theory) is one of the most important challenges in theoretical physics.
Quantum gravity theory is essential for understanding extreme situations, such as the beginning of the Universe.
However, despite great efforts, no one has so far been able to complete the quantum gravity theory.
One of the reasons for the difficulty in unification is that no phenomenon unique to quantum gravity has been experimentally observed because gravitational interactions are weak compared to other fundamental interactions.
Because it is difficult to directly observe the evidence of quantum gravity in experiments using particle accelerators, methods for detecting indirect evidence have been proposed. 
For example, the indirect detection methods using a quantum particle interacting with quantum gravitational waves (gravitons) are discussed~
\cite{Wilczek1, Kanno1, Wilczek2, Wilczek3, Kanno2, Matarrese}.
These studies focused on the loss of quantum coherence of a particle in a quantum superposition state when it interacts with gravitons and discussed the indirect detectability of gravitons.

Thanks to the developments in quantum technology from meso- to macroscopic systems, theoretical proposals have been suggested to directly detect the quantum aspects of gravity in low-energy tabletop experiments
~\cite{Bose2017,Marlleto2017,Christodoulou,Nguyen,Lopez,Blaushi,Miao,Krisnanda,Datta,Matsumura,Miki,Miki2,Miki3,Sugiyama4,Shichijo,Kaku1,Kaku2}.
Authors in Refs.~\cite{Bose2017, Marlleto2017} proposed that two particles each in a quantum superposition state can be entangled states interacting through a Newtonian potential, which is known as the BMV experiment.
This entanglement generation between two particles is understood to be due to the quantum superposition state of the gravitational potentials~\cite{Christodoulou} induced by the particles, which represents the nonclassicality of gravity.
This proposal is expected to be the first step toward experimentally observing the quantum gravity effect.
Motivated by their works, extending models of BMV setup were proposed including the decoherence effect due to the existence of an environment~\cite{Nguyen}, many particles~\cite{Miki}, and internal degrees of freedom~\cite{Kaku1}.
To verify the entanglement generation originating directly from the Newtonian potential, experimental setups focusing on optomechanical systems~\cite{Aspelmeyer} with milligram-scale oscillators have been investigated
~\cite{Miao,Krisnanda,Lopez,Datta,Matsumura,Miki2,Miki3,Sugiyama4,Shichijo,Kaku2}.
However, the meaning of quantum superposition of gravitational potentials and how it is related to the quantization of the gravitational field is not well understood~\cite{Hu, Hall,Eduardo}.

A gedanken experiment for two objects each in a quantum superposition state~\cite{Baym, Mari, Belenchia2018, Belenchia2019, Danielson2021, Sugiyama2, Sugiyama3, Iso1, Iso2} may provide clues to the relation between the meaning of the quantum superposition of gravitational potential and the existence of gravitons.
The inconsistency between the relativistic causality and complementarity proposed in this gedanken experiment is resolved by considering not only the quantum superposition of gravitational fields but also the degrees of freedom of the graviton.
In our previous studies~\cite{Sugiyama2, Sugiyama3}, we investigated the condition that causality and complementary are consistently satisfied in a system where two massive particles are in a superposition state based on the quantum field theory approach.
As a result, we demonstrated that the existence of dynamical degrees of freedom of the gravitational field is a sufficient condition for causality and complementarity to be consistent.
This can be explained as follows. 
Dynamical gravitational fields cause decoherence and suppress the entanglement between two particles.
In particular, the two particles are not entangled when causality is fulfilled because there is no entanglement between them due to decoherence.
Furthermore, the fact that the two particles are not entangled leads to complementarity.
Thus the quantum theory of the gravitational field, in which causality and complementarity are consistent, guarantees the existence of quantum superposition states of the gravitational potential and decoherence between the dynamical gravitational field and a particle.

The decoherence for the state of two particles may be understood as a property of quantum entanglement in multipartite systems.
In general, if tripartite systems A, B, and C are in a pure state, the composite system AB approaches a pure state when the entanglement between A and B becomes stronger.
Then the composite systems AB and C do not correlate.
This property is known as entanglement monogamy~\cite{Coffman, Osborne, Oliveira, Zhu} and is a general property of quantum entanglement.
Thus, in a system in which two particles and a quantized gravitational field interact, if the two particles are always separable, then the composite system consisting of the particle and gravitational field can always be entangled.

In the present paper, we investigate the dynamics of a tripartite system consisting of two quantum particles and a gravitational field, and discuss the structure of entanglement between them.
Through this analysis, we find a trade-off relation (entanglement monogamy) between negativity and conditional von Neumann entropy, which characterizes entanglement.
As a result, if two particles are not entangled with each other, the particle and gravitational field will always be entangled.
Furthermore, we analyze the quantum correlation between the particle in a superposition state and the gravitational field.
This analysis demonstrates that the gravitational field becomes well-superposed if the width of the particle superposition is large.

This paper is organized as follows.
In Sec.~II, we review the results of a QED model, motivated by a two matter-wave interferometer setup.
In Sec.~III, we extend the QED results to a model of a quantized gravitational field and particles.
In Sec.~IV, we discuss the monogamy relation between the negativity and the conditional von Neumann entropy.
Based on this relationship, we derive the condition for entangled states between a particle and a gravitational field.
In Sec.~V, we analyze the behavior of the quantum discord.
Section VI is devoted to the conclusion.
In Appendix A, we briefly introduce the Becchi-Rouet-Stora-Tyutin (BRST) formalism in QED.
In Appendix B, we summarize the QED case formulas and provide a unified description of the gravitational field.
In Appendix C, we present the results of the calculation of $\Gamma^{\text{EM}}_{\text{c}}$, 
$\Phi^{\text{EM}}_{\text{AB}}$, and
$\Phi^{\text{EM}}_{\text{BA}}$ in two specific configurations.
In Appendix D, we derive the two-point function of the quantized gravitational field for the vacuum state.
In Appendix E, we give the proofs of inequality~\eqref{EoF} and Eq.~\eqref{discordin}.
Throughout the present paper, we use the natural units $c=\hbar=1$.

\section{Review of the QED formulation}

\begin{figure}[t]
  \centering
  \includegraphics[width=0.5\linewidth]{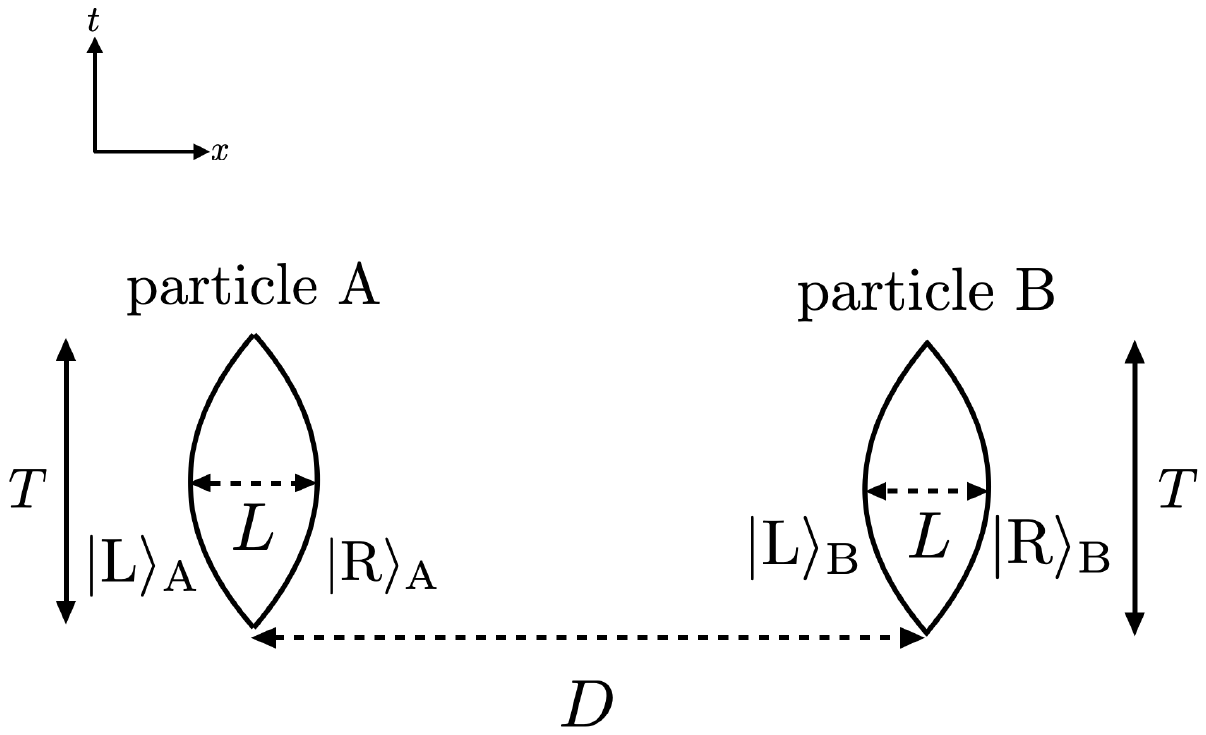}
\caption
{
 Configurations of our model in the regime $T \gg D \gg L$.
 Here $L$ is the length scale of each superposition, $T$ is the coordinate time during which each particle is superposed, and the particles are initially separated by the distance $D$.
 }
  \label{fig:tdlconfiguration}
\end{figure}

Here we review the results of the QED formulation discussed in our previous studies~\cite{Sugiyama1,Sugiyama2,Sugiyama3}. 
This study was motivated by the BMV experimental proposal~\cite{Bose2017,Marlleto2017} for detecting the quantum superposition of the spacetime curvature using two matter-wave interferometers.
This proposal assumes that two massive particles are each in the superposition of two trajectories and interact through a Newtonian potential.
In the next section, we will extend the results of the QED formulation obtained in this section to the case of the gravitational field.
We consider a model of two charged particles, A and B, coupled with an electromagnetic field.
The total Hamiltonian of our system in Schr\"{o}dinger picture is described by the local Hamiltonians of charged particles 
$\hat{H}_{\text{A}}$ and $\hat{H}_{\text{B}}$, the free Hamiltonian of the electromagnetic field 
$\hat{H}_\text{EM}$, and the interaction term $\hat{V}$ as
\begin{eqnarray}
\hat{H}
=
\hat{H}_{\text{A}}+\hat{H}_{\text{B}}+\hat{H}_\text{EM}+\hat{V}, 
\quad 
\hat{V}=\int d^3x \Big(\hat{J}^\mu_{\text{A}}(\bm x)+\hat{J}^\mu_{\text{B}}(\bm{x})\Big)\hat{A}_{\mu}(\bm{x}),
\label{hamiltonian}
\end{eqnarray}
where 
$\hat{J}^\mu_{\text{A}} $ 
and 
$\hat{J}^\mu_{\text{B}} $ are the current operators of each particle coupled with the gauge field operator 
$\hat{A}_\mu $.
Note that the current operator is given by the Dirac field in QED.
The initial state assumes that the two charged particles are in the superposition of two trajectories (Fig.~\ref{fig:tdlconfiguration}) and there is no entanglement between the particles and the electromagnetic field.
Then the initial state of the system is represented by
\begin{align}
|\Psi(0)\rangle
&=
\frac{1}{{2}}
(|\text{L}\rangle_{\text{A}}+|\text{R}\rangle_{\text{A}})
(|\text{L}\rangle_{\text{B}}+|\text{R}\rangle_{\text{B}})
|\alpha\rangle_{\text{EM}},
\label{inistate1}
\end{align}
where
$|\text{L} \rangle_j$ and 
$|\text{R} \rangle_j$
are the localized states of the particles 
$j=\text{A},\text{B}$.
The superposition states of the two charged particles can be realized by considering the Stern-Gerlach effect discussed in~\cite{Bose2017,Marlleto2017}.
This is because the manipulation of an external inhomogeneous magnetic field on a particle with spin degrees of freedom creates a spatially superposed state.
The state  
$|\alpha \rangle_\text{EM}=\hat{D}(\alpha)|0\rangle_\text{EM}$ 
is the initial state of the electromagnetic field. 
$|0\rangle_\text{EM}$ 
is the vacuum state with respect to the electromagnetic field satisfying
$\hat{a}_\mu (\bm{k})|0 \rangle_\text{EM}=0$ 
for the annihilation operator of the electromagnetic field 
$\hat{a}_\mu (\bm{k})$.
The operator
$\hat{D}(\alpha)$ 
is the unitary operator known as a displacement operator, which is defined as
\begin{equation}
\hat{D}(\alpha)
=
\exp
\left[
\int d^3k (\alpha^\mu (\bm{k})\hat{a}^\dagger_\mu (\bm{k})-\text{H.c.})
\right].
\label{D}
\end{equation}
Here the complex function 
$\alpha^\mu (\bm{k})$ 
characterizes the amplitude and phase of initial photon field. 
The form of the complex function 
$\alpha^\mu (\bm{k})$ 
is constrained by the auxiliary condition in the BRST formalism~\cite{Sugiyama1}.
The state 
$|\alpha\rangle_{\text{EM}}$
is interpreted as a coherent state in which a longitudinal mode of the electromagnetic field exists, which follows Gauss's law due to the presence of charged particles (see Ref.~\cite{Sugiyama1}).
In the present analysis, we are only interested in the localized state of each particle, respectively, and the current operator of the field is given by the localized current of each particle,
\begin{align}
\hat{J}^{\mu}_{j\text{I}}(x)
|R\rangle_{j}
\approx
J^{\mu}_{j{\text{R}}}(x)
|R\rangle_{j},
\quad
\hat{J}^{\mu}_{j\text{I}}(x)
|L\rangle_{j}
\approx
J^{\mu}_{j{\text{L}}}(x)
|L\rangle_{j},
\label{approx3}
\end{align}
where 
$\hat{J}^\mu_{i\text{I}}(x)
=e^{i\hat{H}_0 t} \hat{J}^\mu_{i}(\bm{x})e^{-i\hat{H}_0 t}$ 
in the interaction picture with respect to 
$\hat{H}_0 =\hat{H}_{\text{A}} + \hat{H}_{\text{B}} +\hat{H}_\text{EM}$
was introduced.
The explicit forms of
$J^{\mu}_{j{\text{R}}}(x)$ and $J^{\mu}_{j{\text{L}}}(x)$ are
\begin{align}
J_{j\text{R}}^{\mu}(x)
=
e_{j}\int d \tau 
\frac{d X_{j\text{R}}^{\mu}}{d \tau} \delta^{(4)}
\left(x-X_{j\text{R}}(\tau)\right),
\quad 
J_{j\text{L}}^{\mu}(x)
=e_{j} \int d \tau 
\frac{d X_{j\text{L}}^{\mu}}{d \tau} \delta^{(4)}
\left(x-X_{j\text{L}}(\tau)\right),
\end{align}
where 
$X^\mu_{j\text{R}}(\tau)$ and 
$X^\mu_{j\text{L}}(\tau)$ with 
$j=\text{A}, \text{B}$
represent the trajectories of each particle with coupling constants
$e_{\text{A}}$ and $e_{\text{B}}$.
Thus we can proceed with our computation without considering the field degrees of freedom, such as spin. 
In detail, the above equations are valid~\cite{Sugiyama1, Sugiyama2, Sugiyama3, Ford1993, Ford1997, Breuer2001} when the following two conditions are satisfied:
(i) The de Brogile wavelength is smaller than the width of the particle wave packet; 
(ii) The Compton wavelength of the charged particles is much shorter than the wavelength of photons emitted from the charged particles.
The first condition justifies that the state of a particle is localized.
The second condition neglects the processes of a pair creation and annihilation. 
The initial state 
$|\Psi(0)\rangle$ 
evolves as follows:
\begin{align}
|\Psi(T)\rangle 
&= 
\exp
\big[-i \hat{H} T\big]
|\Psi(0)\rangle
\nonumber\\
&=
e^{-i\hat{H}_0T} 
\text{T}\exp
\big[-i\int_{0}^{T} dt \hat{V}_\text{I}(t) \big]
|\Psi(0)\rangle
\nonumber\\
&\approx 
e^{-i\hat{H}_0T}
\frac{1}{2}
\sum_\text{P,Q=R,L}|\text{P}\rangle_{\text{A}} |\text{Q}\rangle_{\text{B}} \hat{U}_\text{PQ} |\alpha\rangle_\text{EM}
\nonumber 
\\
&=
\frac{1}{2}
\sum_\text{P,Q=R,L}|\text{P}_\text{f} \rangle_{\text{A}} |\text{Q}_\text{f} \rangle_{\text{B}} \, e^{-i\hat{H}_\text{EM} T} \hat{U}_\text{PQ} |\alpha\rangle_\text{EM},
\label{state}
\end{align}
where 
$\text{T}$ denotes the time ordering.
We used the approximations provided in~\eqref{approx3} in the third line. 
$|\text{P}_\text{f}\rangle_{\text{A}}
=e^{-i\hat{H}_{\text{A}}T}|\text{P}\rangle_{\text{A}}$ 
and
$|\text{Q}_\text{f}\rangle_{\text{B}}
=e^{-i\hat{H}_{\text{B}}T}|\text{Q}\rangle_{\text{B}}$ 
with $\text{P}, \text{Q}=\text{R}, \text{L}$ 
are the states of charged particles A and B, which moved along the trajectories P and Q, respectively.
The unitary operator 
$\hat{U}_{\text{PQ}}$
is given by:
\begin{align}
\hat{U}_{\text{PQ}} 
=
\text{T} \exp 
\left[
-i \int_{0}^{T} d t \int d^{3} x
\left(J_{\text{AP}}^{\mu}+J_{\text{BQ}}^{\mu}\right) \hat{A}_{\mu}^{\text{I}}(x)
\right],
\end{align}
where  
$\hat{A}^\text{I}_\mu$ 
is the photon field operator in the interaction picture.

Hence, we can explicitly compute the quantum state between the charged particles A and B, respectively.
By tracing out the degrees of freedom of the electromagnetic field, we obtain the reduced density matrix of particles A and B as follows:
\begin{align}
\rho^{\text{EM}}_{\text{AB}} 
&=
\text{Tr}_{\text{EM}}[|\Psi(T)\rangle\langle\Psi(T)|]
\nonumber\\
\quad
&=
\frac{1}{4}
\sum_{\text{P}, \text{Q}=\text{R}, \text{L}}
\sum_{\text{P}', \text{Q}'=\text{R}, \text{L}}
{}_{\text{EM}}\langle \alpha 
|\hat{U}^{\dagger}_{\text{P}'\text{Q}'}\hat{U}_{\text{PQ}}|\alpha \rangle_\text{EM}
\,|\text{P}_\text{f}\rangle_{\text{A}}\langle {\text{P}'}_\text{f}| \otimes
|\text{Q}_\text{f}\rangle_{\text{B}}\langle {\text{Q}'}_\text{f}|
\nonumber\\
\quad
&=
\frac{1}{4}
\sum_{\text{P}, \text{Q}=\text{R}, \text{L}}
\sum_{\text{P}', \text{Q}'=\text{R}, \text{L}}
e^{-\Gamma_{\text{P}'\text{Q}'\text{PQ}}+i\Phi_{\text{P}'\text{Q}'\text{PQ}}}
\,|\text{P}_\text{f}\rangle_{\text{A}}\langle {\text{P}'}_\text{f}| \otimes
|\text{Q}_\text{f}\rangle_{\text{B}}\langle {\text{Q}'}_\text{f}|,
\label{rhoAB}
\end{align}
where the quantities 
$\Gamma_{\text{P}'\text{Q}'\text{PQ}}$ 
and
$\Phi_{\text{P}'\text{Q}'\text{PQ}}$ 
are given by
\begin{align}
\Gamma_{\text{P}'\text{Q}'\text{PQ}}
&=
\frac{1}{4}
\int d^4x \int d^4y 
(J^\mu_{\text{P}'\text{Q}'}(x)-J^\mu_{\text{PQ}}(x))
(J^\nu_{\text{P}'\text{Q}'}(y)-J^\nu_{\text{PQ}}(y))
\langle\bigl\{
\hat{A}^\text{I}_\mu (x), \hat{A}^\text{I}_\nu (y)
\bigr\}\rangle,
\\
\Phi_{\text{P}'\text{Q}'\text{PQ}}
&=
\int d^4x 
(J^\mu_{\text{P}'\text{Q}'}(x)-J^\mu_{\text{PQ}}(x))A_\mu (x)
\nonumber\\
\quad
&
-\frac{1}{2}
\int d^4x \int d^4y 
(J^\mu_{\text{P}'\text{Q}'}(x)-J^\mu_{\text{PQ}}(x))(J^\nu_{\text{P}'\text{Q}'}(y)+J^\nu_{\text{PQ}}(y))
G^\text{r}_{\mu \nu}(x,y).
\label{PhiP'Q'PQ}
\end{align}
Here 
$\langle \bigl\{
\hat{A}^\text{I}_{\mu}(x), \hat{A}^\text{I}_{\mu}(y)
\bigr\}\rangle$ 
and 
$G^{\text{r}}_{\mu\nu}(x,y)$ 
are the two-point function of the vacuum state 
$|0\rangle_{\text{EM}}$ 
and the retarded Green's function with respect to the gauge field in the interaction picture introduced as
\begin{align}
\langle\bigl\{
\hat{A}^\text{I}_{\mu}(x), \hat{A}^\text{I}_{\mu}(y)
\bigr\}\rangle
=
\frac{\eta_{\mu \nu}}{4 \pi^{2}}
\left(
\frac{1}{-\left(x^{0}-y^{0}-i \epsilon\right)^{2}+|\boldsymbol{x}-\boldsymbol{y}|^{2}}
+\text{c.c.}
\right)
\end{align} 
with the UV cutoff parameter $\epsilon$ and
\begin{align}
G^\text{r}_{\mu \nu} (x,y)
=
-i[\hat{A}^\text{I}_\mu (x), \hat{A}^\text{I}_\nu (y)]
\theta (x^0-y^0). 
\end{align}
Field $A_\mu (x)$ in~\eqref{PhiP'Q'PQ} is the coherent electromagnetic field, which is
\begin{equation}
A_{\mu} (x)
= \int \frac{d^3k}{(2\pi)^{3/2} \sqrt{2k^0}}
\left(
\alpha_\mu (\bm{k})e^{ik_\nu x^\nu}+\text{c.c.}
\right),
\end{equation}
and the complex function 
$\alpha_{\mu}(\bm{k})$
satisfies 
\begin{equation}
k^{\mu} \alpha_{\mu}(\bm{k})
=
-\frac{\tilde{J}^{0}(\bm{k})}{\sqrt{2k^{0}}}
\label{brstcondition}
\end{equation}
to ensure the BRST condition (for a more detailed discussion see Appendix A and Ref.\cite{Sugiyama1}).  
$\tilde{J}^0(\bm{k})
=
\tilde{J}^0_\text{A}(\bm{k})+\tilde{J}^0_\text{B}(\bm{k})$
is the eigenvalue of the Fourier transform of the charged current
$\hat{\tilde{J}}^0(\bm{k})
=
\hat{\tilde{J}}^0_\text{A}(\bm{k})+\hat{\tilde{J}}^0_\text{B}(\bm{k})$ at the initial time 
$t=0$.
The quantum states of particle A (B) is also obtained by tracing out the degrees of freedom of particle B (A) of the density matrix 
$\rho^{\text{EM}}_{\text{AB}}$,
\begin{align}
\rho^{\text{EM}}_{\text{A}}
&=
\text{Tr}_{\text{B}, \text{ph}}[|\Psi(T)\rangle \langle\Psi(T)|]
\nonumber\\
\quad
&=
\frac{1}{2}
\begin{pmatrix}
1&\frac{1}{2}e^{-\Gamma^{\text{EM}}_{\text{A}}+i\Phi^{\text{EM}}_\text{A}}\Big(e^{-i\int d^4x \Delta J^{\mu}_{\text{A}}(x) A_{\text{BR} \mu}(x)}+e^{-i\int d^4x\Delta J^{\mu}_{\text{A}}(x) A_{\text{BL} \mu}(x)}\Big)
\\
\quad * \quad &1
\end{pmatrix}
\label{rhoA},
\end{align}
and 
\begin{align}
\rho^{\text{EM}}_{\text{B}}
&=
\text{Tr}_{\text{A}, \text{ph}}[|\Psi(T)\rangle \langle\Psi(T)|]
\nonumber\\
\quad
&=
\frac{1}{2}
\begin{pmatrix}
1&\frac{1}{2}e^{-\Gamma^{\text{EM}}_{\text{B}}+i\Phi^{\text{EM}}_\text{B}}\Big(e^{-i\int d^4x\Delta J^{\mu}_{\text{B}}(x) A_{\text{AR} \mu}(x)}+e^{-i\int d^4x\Delta J^{\mu}_{\text{B}}(x) A_{\text{AL} \mu}(x)}\Big)
\\
\quad * \quad &1
\end{pmatrix}
\label{rhoB},
\end{align}
where we defined 
$\Delta J^\mu_i= J^\mu_{i\text{R}}- J^\mu_{i\text{L}}$
and used the basis 
$\{|\text{R}_\text{f} 
\rangle_\text{A}, |\text{L}_\text{f} \rangle_\text{A} \}$ 
to represent the density operator.
Note that the symbol $*$ shows the complex conjugate of the 
$(\text{R},\text{L})$ 
off-diagonal component. 
Function 
$A^{\mu}_{i\text{P}} (x)$
with $i=\text{A}, \text{B}$ 
and $\text{P}=\text{R}, \text{L}$ 
corresponds to the retarded potentials defined by
\begin{align}
A^{\mu}_{i \text{P}}(x)
=
\int d^{4} y 
G_{\mu \nu}^{r}(x, y) J_{i \text{P}}^{\nu}(y).
\end{align}
Here the quantities
$\Gamma^{\text{EM}}_i$ $(i=\text{A}, \text{B})$,
$\Phi^{\text{EM}}_\text{A}$ and
$\Phi^{\text{EM}}_\text{B}$
are introduced as
\begin{align}
\Gamma^{\text{EM}}_{i}
&=
\frac{1}{4}
\int d^4x\int d^4y 
\Delta J^{\mu}_i(x) \Delta J^{\nu}_i(y)
\langle \{\hat{A}^\text{I}_{\mu}(x), \hat{A}^\text{I}_{\nu}(y)\}\rangle,
\label{gammaiEM}
\\
\Phi^{\text{EM}}_\text{A}
&=
\int d^4x 
\Delta J^{\mu}_{\text{A}}(x) A_{\mu}(x)
-\frac{1}{2}
\int d^4x d^4y
\Delta J^{\mu}_{\text{A}}(x) 
(J^{\nu}_{\text{AR}}(y)+J^{\nu}_{\text{AL}}(y))
G^{\text{r}}_{\mu\nu}(x,y),
\\
\Phi^{\text{EM}}_\text{B}
&=
\int d^4x 
\Delta J^{\mu}_{\text{B}}(x)A_{\mu}(x)
-\frac{1}{2}
\int d^4x d^4y
\Delta J^{\mu}_{\text{B}}(x)
(J^{\nu}_{\text{BR}}(y)+J^{\nu}_{\text{BL}}(y))
G^{\text{r}}_{\mu\nu}(x,y).
\end{align}
The density matrix for the electromagnetic fields in Eq.~\eqref{rhoA} and~\eqref{rhoB} is convenient for calculating the von Neumann entropy in the case of an electromagnetic field with respect to quantum system A (B)~[Eq.~\eqref{valueAB}].
Similarly, the density matrix in Eq.~\eqref{rhoAB} is used to compute the von Neumann entropy for the composite system AB 
[Eqs.~\eqref{lambdam} and~\eqref{lambdap}], 
and negativity [Eq.~\eqref{negativityEM}], 
and the concurrence between two massive particles, A and B 
[Eqs.~\eqref{alpha1}-\eqref{alpha4}].
As explained in Appendix A, 
the von Neumann entropy, the negativity, and the concurrence are described by 
$\Gamma^{\text{EM}}_{i}$ [Eq.~\eqref{gammaiEM}], 
$\Gamma^{\text{EM}}_{\text{c}}$ [Eq.~\eqref{gammacEM}], 
$\Phi^{\text{EM}}_{\text{AB}}$, 
and 
$\Phi^{\text{EM}}_{\text{BA}}$ [Eq~\eqref{phiabEM}].
In the next section, we extend the quantities
$\Gamma^{\text{EM}}_{i}$, 
$\Gamma^{\text{EM}}_{\text{c}}$, 
$\Phi^{\text{EM}}_{\text{AB}}$, 
and 
$\Phi^{\text{EM}}_{\text{BA}}$
to the gravitational version using the unified description introduced in Appendix A.

\section{Setup of massive particles with gravitational field}
In this section, we consider a linearized gravity theory coupled with two massive particles A and B, whose masses are $m$.
In our analysis, we treat the gravitational field as quantized and each particle is in a spatially localized superposition state.
The two particles are initially separated by a distance $D$ and maintain a spatially superposed state with the separation $L$ during the time $T$.
We assume that the two particles have a nonrelativistic motion.
\begin{figure}[H]
  \centering
  \includegraphics[width=0.5\linewidth]{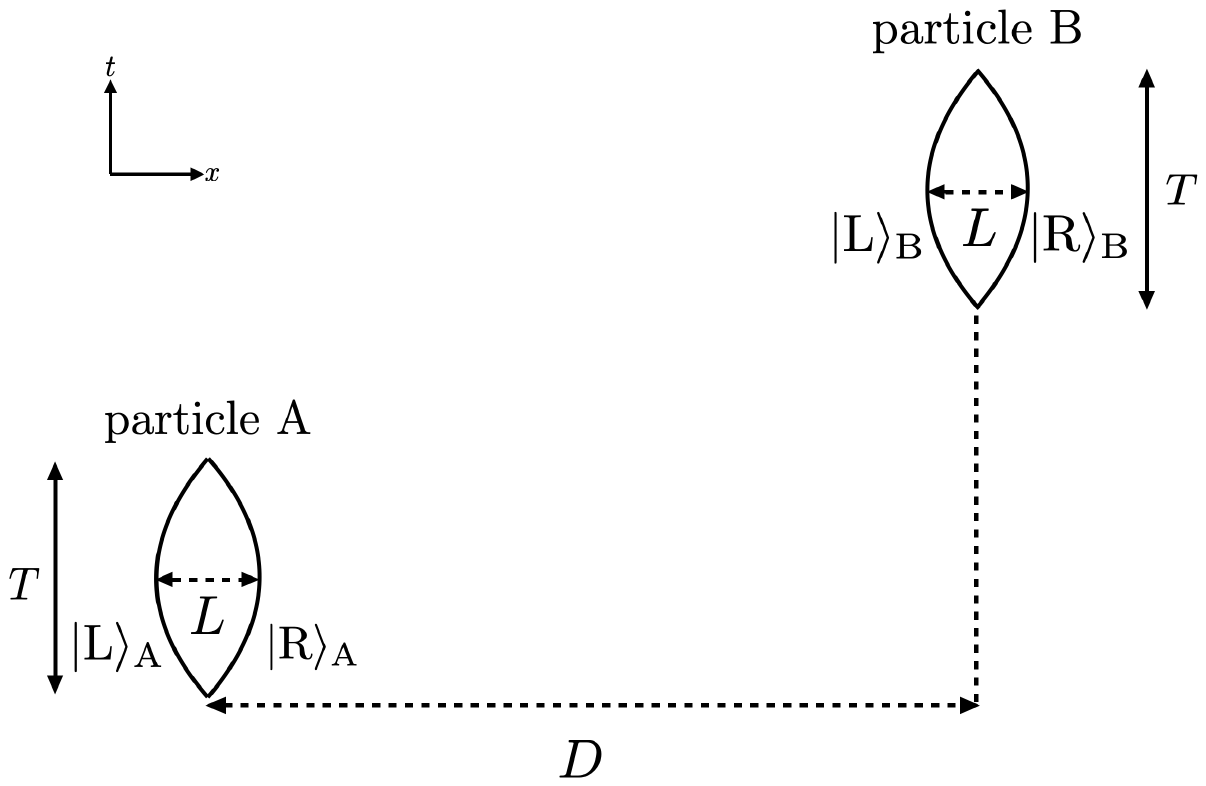}
\caption
{
 Configurations of our model in the regime $D \gg T \gg L$.
 In this regime, the retarded Green's function propagating from particle B to A vanishes.
 }
  \label{fig:dtlconfiguration}
\end{figure}
In particular, as shown in Figs.~\ref{fig:tdlconfiguration} and~\ref{fig:dtlconfiguration}, we discuss two types of configurations; one configuration can send information to each other, but the other cannot.
The gravitational field is treated as a linearized gravity by expanding the metric of spacetime 
$g_{\mu\nu}$ 
around Minkowski spacetime background 
$\eta_{\mu\nu}$:
$g_{\mu\nu}=\eta_{\mu\nu}+h_{\mu\nu}$, where 
$h_{\mu\nu}$ 
is the metric perturbation satisfying 
$|h_{\mu\nu}|\ll1$.
\footnote{
Based on the linearized gravity theory~\cite{Wheeler, Sugiyama2}, energy-momentum tensor of a particle $T_{\mu\nu}$ induces the fluctuation part of the metric 
$h_{\mu\nu}
\sim
G\int d^3y {T_{\mu\nu}(t_{\text{r}}, \bm{y})}/{|\bm{x}-\bm{y}|}.$
Here 
$t_{\text{r}}=t-|\bm{x}-\bm{y}|$
is the retarded time, which represents the delay with respect to propagation from the source point $\bm{y}$ to a spacetime point $\bm{x}$.
The components of $h_{\mu\nu}$ are evaluated as
$
h_{00} \sim G{m}/{R},
\quad
h_{0i} \sim h_{00} ({L}/{T})\bm{e}_{i},
\quad
h_{ij} \sim h_{00} ({L}/{T})^2\bm{e}_{i}\bm{e}_{j},
$
where $R$ characterizes the typical length scale of particles A and B satisfying 
$R \lesssim L$ 
(for a detail, see Chapter 36.10 of Ref.~\cite{Wheeler}).
$(L/T)\bm{e}_{i}$ denotes the characteristic velocity of the system in the $i$ direction of the unit vector $\bm{e}_{i}$.
We regard the length scale $R$ as the typical size of the particle.
Considering the nonrelativistic condition $L/T \ll 1$, the condition $|h_{\mu\nu}|\ll1$ is valid when
$
1
\gg
G{m}/{L}
=
{g^2}/{mL}
=
g^2{\lambda_{\text{C}}}/{L}
\label{condition}
$
is satisfied.
Here we introduced $g^2=Gm^2$ of the coupling constant between the two particles with the gravitational constant $G$.
$\lambda_{\text{C}}=1/m$ 
is the Compton wave length of the two particles.
In the following analysis, we choose the parameter with which the above condition~\eqref{condition} is satisfied.
}
In particular, we focus on the similarity between the electromagnetic field and gravitational fields, and extend the results obtained from the analysis of an electromagnetic field in Sec. II to the gravitational field (see also our previous study~\cite{Sugiyama2}).
In this extension, the current of the charged particle 
$J^{\mu}_{\text{AP}}(x)$ ($J^{\mu}_{\text{BQ}}(x)$) 
is replaced as the energy-momentum tensor of particle A (B) as 
${T}^{\mu \nu}_{\text{AP}}(x)$ (${T}^{\mu \nu}_{\text{BQ}}(x)$) 
localized around their trajectories of P (Q)$(=R, L)$.
Using the unified notation introduced in Appendix A, the decoherence and the entanglement between the two massive particles can be described by Eqs.~\eqref{gammai}, \eqref{gammac}, and \eqref{phi}
\begin{align}
\Gamma_{i}
&=
\frac{1}{4}
\int d^4x\int d^4y 
\Delta T^{\mu\nu}_{i}(x)\Delta T^{\rho\sigma}_{i}(y)
\langle 
\{\hat{h}^\text{I}_{\mu\nu}(x), \hat{h}^\text{I}_{\rho\sigma}(y)\}
\rangle,
\label{gammai}
\\
\quad
\Gamma_\text{c}
&=
\frac{1}{2}
\int d^4x\int d^4y
\Delta T^{\mu\nu}_{\text{A}}(x)\Delta T^{\rho\sigma}_{\text{B}}(y)
\langle 
\{\hat{h}^\text{I}_{\mu\nu}(x), \hat{h}^\text{I}_{\rho\sigma}(y)\}
\rangle,
\label{gammac}
\end{align}
where we defined 
$\Delta T^{\mu\nu}_{i}
=T^{\mu\nu}_{i\text{R}}-T^{\mu\nu}_{i\text{L}}$ 
with $i=\text{A}, \text{B}$.
$\langle\bigl\{
\hat{h}^\text{I}_{\mu\nu} (x), \hat{h}^\text{I}_{\rho\sigma} (y)
\bigr\}\rangle$
is the two-point function of the vacuum state $|0\rangle$ with respect to $\hat{h}^{\text{I}}_{\mu\nu}(x)$ in the interaction picture. 
The quantities $\Phi_{\text{AB}}$ and $\Phi_{\text{BA}}$ are introduced as:
\begin{align}
\Phi_{\text{AB}}
&=
\int d^4x d^4y
\Delta T^{\mu\nu}_{\text{A}}(x)\Delta T^{\rho\sigma}_{\text{B}}(y)
G^{\text{r}}_{\mu\nu\rho\sigma}(x,y),
\quad
\Phi_{\text{BA}}
=
\int d^4x d^4y
\Delta T^{\mu\nu}_{\text{B}}(x)\Delta T^{\rho\sigma}_{\text{A}}(y)
G^{\text{r}}_{\mu\nu\rho\sigma}(x,y)
\label{phi}
\end{align}
with the gravitational version of retarded Green's function $G^\text{r}_{\mu \nu \rho \sigma} (x,y)$ 
(for details, see~\cite{Donoghue}).

Because we analyze quantitatively, we evaluate the quantities 
$\Gamma_{i}$, $\Gamma_{\text{c}}$, $\Phi_{\text{AB}}$, 
and $\Phi_{\text{BA}}$ 
by order estimation up to the numerical factors.
The quantities $\Gamma_{\text{A}}$ and $\Gamma_{\text{B}}$ are estimated as the number of gravitons emitted by the quadrupole radiation during time $T$ per the energy of a single graviton
\footnote{
The total power of the gravitons $W$ emitted by quadrupole radiation during time $T$ is evaluated as
$W\sim G(Q/T^3)^2T=g^2(L^4/T^5)$ 
with the mass quadrupole 
$Q=mL^2$ 
(for a more detailed explanation, see Chapter 36.1 and 36.2 in~\cite{Wheeler}).
The energy of a single graviton $\nu$ is $\nu=1/T$ in the unit 
$\hbar=1$, and we obtain the total number of gravitons 
$W/\nu\sim g^2(L/T)^4$, which is consistent with the result of Ref.
~\cite{Belenchia2018}.
}
\begin{align}
\Gamma_{\text{A}}
=
\Gamma_{\text{B}}
\approx
g^2
\left(\frac{L}{T}\right)^4
,
\label{gammaig}
\end{align}
where we introduced $g^2=Gm^2$ of the coupling constant between the two particles with the gravitational constant $G$, whereas the electromagnetic case of the quantities 
$\Gamma^{\text{EM}}_{\text{A}}$ and $\Gamma^{\text{EM}}_{\text{B}}$, which corresponds to the dipole radiation, are~\eqref{gammaiem}
\begin{align}
\Gamma^{\text{EM}}_{\text{A}}
=
\Gamma^{\text{EM}}_{\text{B}}
\approx
e^2
\left(\frac{L}{T}\right)^2
\end{align}
with the electric charges $e=e_A=e_B$.
Note that the quantities 
$\Gamma_{\text{A}}$ and $\Gamma_{\text{B}}$ 
are determined by the parameters of their system A and B, respectively.
In contrast, the quantities 
$\Phi_{\text{AB}}$, $\Phi_{\text{BA}}$, 
and $\Gamma_{\text{c}}$ 
characterize the correlation between two particles.
Thus the distance between systems A and B is important.
In the regime $D \gg T \gg L$, the quantities $\Phi_{\text{AB}}$, $\Phi_{\text{BA}}$, and $\Gamma_{\text{c}}$ will be given by
\footnote{
Since the quantities 
$\int d^3x \Delta T^{\mu\nu}_{i} \sim m(L/T)^2$ 
with 
$i=\text{A}, \text{B}$ 
represent the energy of the particle, the quantity 
$\Gamma_{\text{c}}$ 
is estimated together with the two-point function Eq.~\eqref{twog} to be Eq.~\eqref{dtl}.
Note that $(T/D)^4$ was obtained by analogy with the result of the electromagnetic field~\eqref{eq:GammacA1}.
The effect of $\Gamma_{\text{c}}$ is so small in the region $D \gg T \gg L$~\cite{Sugiyama1} that it is expected not to affect entanglement between two particles.
More accurate calculations of the quantities 
$\Gamma_{\text{A}}$, $\Gamma_{\text{B}}$, and $\Gamma_{\text{c}}$
should include the energy-momentum tensor of the Stern-Gerlach apparatus due to the conservation of the energy-momentum tensor~\cite{Suzuki}.
}
\begin{align}
\Phi_{\text{AB}}
=0,
\quad
\Phi_{\text{BA}}
\approx
g^2
\left(\frac{L}{T}\right)^2\left(\frac{T}{D}\right)^3,
\quad
|\Gamma_{\text{c}}|
\approx
g^2\left(\frac{L}{T}\right)^4\left(\frac{T}{D}\right)^4,
\label{dtl}
\end{align}
where $\Phi_{\text{AB}}=0$ is understood as the vanishing of the retarded Green's function propagating from the particle B to A~\cite{Sugiyama1, Sugiyama2, Sugiyama3}.
The quantity $\Phi_{\text{BA}}$ is the phase induced by the Newtonian potential between the massive particles.
\footnote{
The phase $\Phi_{\text{BA}}$ 
is estimated as the difference of relative phase shift 
$\phi_{\text{R}}$ and $\phi_{\text{L}}$ 
due to gravitational potential induced by particle A in the superposition state during time $T$.
Here $\phi_{\text{R}}$ ($\phi_{\text{L}}$) is the relative phase perceived by superposed particle B when particle A goes through the right (left) side of the path. 
Thus $\Phi_{\text{BA}}$ is computed as 
$\Phi_{\text{BA}}=\phi_{\text{R}}-\phi_{\text{L}}
\approx
-Gm^2T(1/(D-L)-1/D)+Gm^2T(1/D-1/(D+L))
\approx g^2(L/T)^2(T/D)^3$, where we neglected the numerical factors.
}
$\Gamma_{\text{c}}$ is referred to the result of the order estimation~\eqref{eq:GammacA1} presented in Appendix~\ref{QED}.
On the contrary, in the regime $T \gg D \gg L$, the quantities $\Phi_{\text{AB}}$, $\Phi_{\text{BA}}$, and $\Gamma_{\text{c}}$ will be of order
\begin{align}
\Phi_{\text{AB}}
=
\Phi_{\text{BA}}
\approx
g^2
\left(\frac{L}{T}\right)^2\left(\frac{T}{D}\right)^3,
\quad
|\Gamma_{\text{c}}|
\approx
g^2\left(\frac{L}{T}\right)^4.
\label{tdl}
\end{align}
Note that, in the regime $T \gg D \gg L$, $\Phi_{\text{AB}}$ can be equivalent to $\Phi_{\text{BA}}$ because of the symmetric configuration of the systems A and B. 
The quantity $\Gamma_{\text{c}}$ is estimated by using Eq.~\eqref{gammactdl} in Appendix~\ref{QED}, where we ignored the term proportional to $D/T$ because of $D/T \ll 1$.
In the following two sections, we consider the quantumness of the gravitational field in terms of entanglement monogamy.
Entanglement monogamy refers to the feature that entanglement cannot be freely shared among multiple parties.
To analyze the quantumness of the gravitational field quantitatively, we investigate the behavior of the entanglement of formation (in Sec. IV) and the quantum discord (in Sec. V) between the massive particle A and the gravitational field.

\section{Quantumness of gravitational field due to monogamy relation}

In this section, we discuss the entanglement between particle A and the gravitational field from the viewpoint of entanglement monogamy.
To judge whether particle A and the gravitational field are entangled or not, we consider the entanglement of formation 
$E_{\text{f}}(\rho_{\text{A,g}})$.
The entanglement of formation is one of the quantities that determines whether two quantum systems are in an entangled state or not.
For example, if $E_{\text{f}}(\rho_{\text{A,g}})>0$, then particle A and the gravitational field are entangled.
However, if the entanglement of formation vanishes: 
$E_{\text{f}}(\rho_{\text{A,g}})=0$, particle A and the gravitational field are not entangled.
The entanglement of formation has a lower bound due to the conditional von Neumann entropy $S({\text{A}|\text{B}})$~\cite{Cerf};
\begin{align}
E_{\text{f}}(\rho_{\text{A,g}})
\geq
S(\text{A}|\text{B}),
\label{EoF}
\end{align}
where the proof of this inequality is presented in Appendix~\ref{proofEoF}.
Thus inequality~\eqref{EoF} indicates that particle A and the gravitational field are entangled when $S({\text{A}|\text{B}})>0$.
The conditional von Neumann entropy $S({\text{A}|\text{B}})$ is analogous to the classical conditional entropy as
\begin{align}
S({\text{A}|\text{B}})
:=
S(\rho_{\text{AB}})-S(\rho_{\text{B}}).
\label{conditionalE}
\end{align}
The von Neumann entropy $S(\rho_{\text{X}})$ measures how strong the correlation is between subsystem $X$ and its complement system $\bar{X}$.
In classical theory, the conditional entropy is always positive; 
however, in quantum theory, it can be negative~\cite{Horodecki}.
The negativity of the conditional von Neumann entropy, $S({\text{A}|\text{B}})$, can be roughly interpreted as the entanglement between the two systems A and B.
The von Neumann entropy $S(\rho_{\text{B}})$ is computed as follows:
\begin{align}
S(\rho_{\text{B}})
&=
-\sum_{s=\pm}
\lambda_{\text{s}}[\rho_{\text{B}}]
\log[\lambda_{\text{s}}[\rho_{\text{B}}]],
\quad
\lambda_{\pm}[\rho_{\text{B}}]
=
\frac{1}{2}
\left[
1\pm e^{-\Gamma_{\text{B}}}\cos
\Big(
\frac{\Phi_{\text{BA}}}{2}
\Big)
\right],
\end{align}
where the eigenvalues $\lambda_{\pm}[\rho_{\text{A}}]$ 
are obtained by extending the quantities in Eqs.~\eqref{valueAB} to a gravitational version.
The von Neumann entropy $S(\rho_{\text{AB}})$ is derived as 
\begin{align}
S(\rho_{\text{AB}})
&=
-\sum_{s=\pm}
\Big(
\lambda^{\text{s}}_{1}[\rho_{\text{AB}}]\log[\lambda^{\text{s}}_{1}[\rho_{\text{AB}}]]
+
\lambda^{\text{s}}_{2}[\rho_{\text{AB}}]\log[\lambda^{\text{s}}_{2}[\rho_{\text{AB}}]]
\Big)
\end{align}
with the eigenvalues of the density matrix
$\rho_{\text{AB}}$ obtained from Eqs.~\eqref{lambdam} and \eqref{lambdap}
\begin{align}
\lambda^{\pm}_{1}[\rho_{\text{AB}}]
&=
\frac{1}{4}
\left[
1-e^{-\Gamma_\text{A}-\Gamma_\text{B}} 
\cosh[\Gamma_\text{c}]
\pm 
\Big\{
\big(
e^{-\Gamma_{\text{A}}}-e^{-\Gamma_{\text{B}}}
\big)^2
+4e^{-\Gamma_{\text{A}}-\Gamma_{\text{B}}}
\sin^2\Big[\frac{\Phi_{\text{AB}}-\Phi_{\text{BA}}}{4}\Big]
+e^{-2\Gamma_\text{A}-2\Gamma_\text{B}} \sinh^2[\Gamma_\text{c}]
\Big\}^{\frac{1}{2}}
\right],
\label{eigenSABm}
\\
\quad
\lambda^{\pm}_{2}[\rho_{\text{AB}}]
&=
\frac{1}{4}
\left[
1+e^{-\Gamma_\text{A}-\Gamma_\text{B}}\cosh[\Gamma_\text{c}]
\pm 
\Big\{
\big(
e^{-\Gamma_{\text{A}}}-e^{-\Gamma_{\text{B}}}
\big)^2
+4e^{-\Gamma_{\text{A}}-\Gamma_{\text{B}}}
\cos^2\Big[\frac{\Phi_{\text{AB}}-\Phi_{\text{BA}}}{4}\Big]
+e^{-2\Gamma_\text{A}-2\Gamma_\text{B}} \sinh^2[\Gamma_\text{c}]
\Big\}^{\frac{1}{2}}
\right].
\label{eigenSABp}
\end{align}
In the following analysis, we evaluate the conditional von Neumann entropy 
$S(\text{A}|\text{B})$
in two regimes 
$D \gg T \gg L$ and $T \gg D \gg L$.
\subsection{$D \gg T \gg  L$ regime}
\begin{figure}[t]
  \centering
  \begin{minipage}[b]{0.4\linewidth}
  \includegraphics[width=1\linewidth]{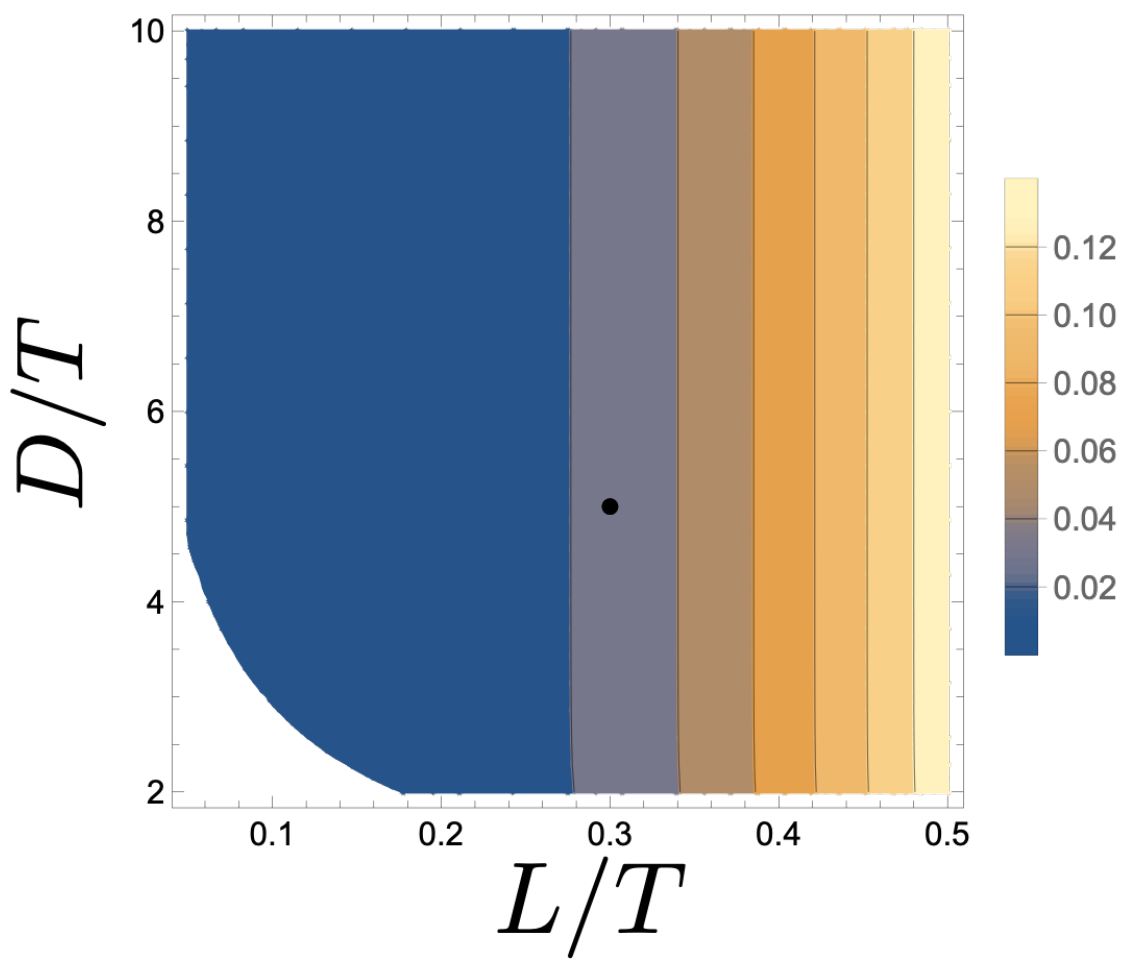}
  \end{minipage}
  \begin{minipage}[b]{0.4\linewidth}
  \includegraphics[width=1\linewidth]{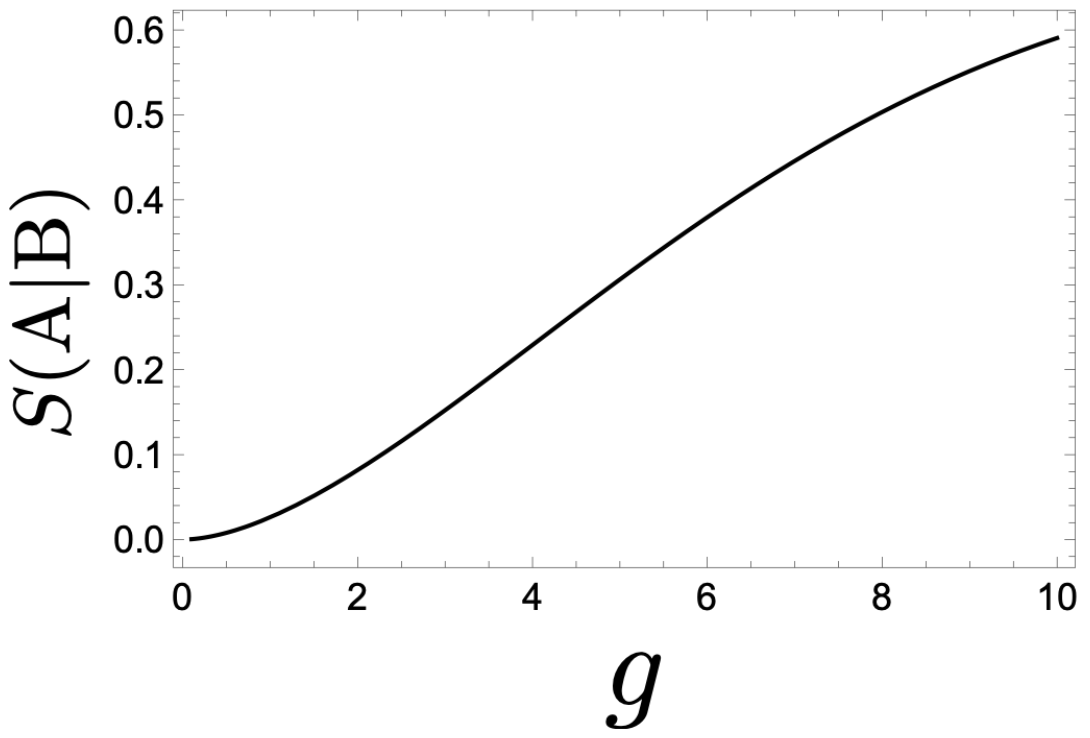}
  \end{minipage}
  \caption{
  Left panel: Contour plots of the conditional von Neumann entropy $S({\text{A}|\text{B}})$ as functions of $L/T$ and $D/T$, where we adopted the coupling constant $g=1$. 
  The black circle is a point when $D/T=5$ with $L/T=3/10$.
  Right panel: Conditional von Neumann entropy $S({\text{A}|\text{B}})$ as a function of coupling constant $g$.
  This graph assumes the parameters of the black circle in the left panel.
  \label{fig:condE}}
\end{figure}
The left panel of Fig.~\ref{fig:condE} shows the parameters dependence of the conditional von Neumann entropy $S(\text{A}|\text{B})$.
To obtain a qualitative understanding of the behavior of the left panel in Fig.~\ref{fig:condE}, we approximate the conditional von Neumann entropy as
\begin{align}
S({\text{A}|\text{B}})
\approx
\frac{\Gamma_{\text{B}}}{2}
\left(
1-\log\left[\frac{\Gamma_{\text{B}}}{2}\right]
\right)
\label{SABapp},
\end{align}
where we used 
$\Gamma_{\text{A}}=\Gamma_{\text{B}} \ll 1$ and $\Phi_{\text{BA}}\ll1$, and we assumed the condition $\Phi_{\text{BA}}/4\Gamma_{\text{B}}\ll1$.
The above Eq.~\eqref{SABapp} is independent of the quantity of $\Phi_{\text{BA}}$, that is, $D/T$ and its amount depends only on 
$\Gamma_{\text{B}}=g^2(L/T)^4$.
This figure represents that $S(\text{A}|\text{B})$ is always positive and does not depend on the distance between two particles.
The independence of the distance $D$ can be understood by introducing an entanglement measure called negativity.
The negativity $\mathcal{N}_{\text{AB}}$ characterizes the entanglement between two particles~\cite{Vidal2002,Sanpera1998}.
In particular, two particles A and B are regarded as the two-qubit in our system, and then the negativity is given as follows:
\begin{align}
\mathcal{N}_{\text{AB}}
=\max[-\lambda_\text{min},0]
\end{align}
with the minimum eigenvalue $\lambda_\text{min}$
\begin{align}
\lambda_{\text{min}}
&=
\frac{1}{4}
\Big[
1-e^{-\Gamma_{\text{A}}-\Gamma_{\text{B}}} \cosh[\Gamma_\text{c}]- 
\Big\{
\big(e^{-\Gamma_{\text{A}}}-e^{-\Gamma_{\text{B}}}\big)^2+4e^{-\Gamma_{\text{A}}-\Gamma_{\text{B}}}\sin^2\Big[\frac{\Phi_{\text{AB}}+\Phi_{\text{BA}}}{4}\Big]+e^{-2\Gamma_{\text{A}}-2\Gamma_{\text{B}}} \sinh^2[\Gamma_\text{c}]
\Big
\}^{\frac{1}{2}}
\Big],
\label{negativity}
\end{align}
where the unified notation was applied.
If 
$\mathcal{N}_{\text{AB}}=0$ or $\lambda_{\text{min}} \geq 0$ 
holds, two particles are not entangled.
In our previous studies~\cite{Sugiyama1,Sugiyama3}, we pointed out that negativity $\mathcal{N}_{\text{AB}}$ vanishes in the regime $D \gg T \gg L$ because of the existence of the vacuum fluctuations 
$\Gamma_{\text{A}}$ and $\Gamma_{\text{B}}$.
Thus, there is no entanglement between A and B.
Note that the white region in Fig.~\ref{fig:condE} may suggest that the approximation to derive the quantities 
$\Gamma_{\text{A}}$, $\Gamma_{\text{B}}$, $\Gamma_{\text{c}}$, 
and $\Phi_{\text{BA}}$ is invalid. 

The right panel of Fig.~\ref{fig:condE} shows the behavior of $S(\text{A}|\text{B})$ versus the coupling constant $g$, respectively.
In the limit of $g\rightarrow0$ there is no interaction among particle A, B, and the gravitational field; therefore, the quantum state $\rho_{\text{AB}}$ and its reduced density matrix $\rho_{\text{B}}$ become pure state, i.e., 
$S(\rho_{\text{AB}})=S(\rho_{\text{B}})=0$.
In contrast, in the limit of $g\rightarrow \infty$, the decoherences
$\Gamma_{\text{A}}$ and $\Gamma_{\text{B}}$ 
are dominant, and then the quantum states
$\rho_{\text{AB}}$ and $\rho_{\text{B}}$
approaches the classical mixed state
\begin{align}
\rho_{\text{AB}}
\rightarrow
\frac{1}{4}\mathbb{1}_{4\times4},
\quad
\rho_{\text{B}}
\rightarrow
\frac{1}{2}\mathbb{1}_{2\times2}
\end{align}
with $n\times n$ identity matrix $\mathbb{1}_{n\times n}$.
These limits lead to $S(\text{A}|\text{B}) \to \log2$ for $g \to \infty$.
Thus, in the region $D \gg T \gg L$, the conditional von Neumann entropy $S(\text{A}|\text{B})$ is always positive.
Therefore, $E_{\text{f}}(\rho_{\text{A,g}})>0$ is constantly fulfilled because of the inequality~\eqref{EoF}.

The condition $E_{\text{f}}(\rho_{\text{A,g}})>0$ can also be understood from the viewpoint of the monogamy relation.
$\lambda_{\text{min}}\geq0$ gives the concrete reason of the positivity of the conditional von Neumann entropy as
\begin{align}
0=\mathcal{N}_{\text{AB}}
=
\lambda_{\text{min}}\geq0
\quad
\Rightarrow
\quad
0=E_{\text{f}}(\rho_{\text{AB}})
\geq
-S(\text{A}|\text{B}),
\label{posSAB}
\end{align}
where the inequality~\eqref{ABE} was used in the right-hand side of 
Eq.~\eqref{posSAB}.
Note that, in general, 
$\mathcal{N}_{\text{AB}}>0$ 
is equivalent to $E_{\text{f}}(\rho_{\text{AB}})>0$ 
if the composite system AB is a two-qubit system, which leads to 
$\mathcal{N}_{\text{AB}}=0 
\Leftrightarrow E_{\text{f}}(\rho_{\text{AB}})=0$ 
due to the contraposition.
By combining the inequality~\eqref{EoF} and the above relation~\eqref{posSAB}, we obtain the following result:
\begin{align}
0=\mathcal{N}_{\text{AB}}
=
\lambda_{\text{min}}
\geq
0
\quad
\Rightarrow
\quad
E_{\text{f}}(\rho_{\text{AB}})=0
\quad
\Rightarrow
\quad
S({\text{A}|\text{B}})>0
\quad
\Rightarrow
\quad
E_{\text{f}}(\rho_{\text{A,g}}) > 0,
\label{result}
\end{align}
where 
$S({\text{A}|\text{B}})>0$ is satisfied in the regime 
$D \gg T \gg L$.
This implies that particle A and gravitational field are always entangled when two particles A and B are not entangled.
Thus, this behavior indicates a monogamy relation among particle A, B, and the gravitational field.

\subsection{$T \gg D \gg L$ regime}
\begin{figure}[t]
 \centering
  \begin{minipage}[b]{0.4\linewidth}
  \includegraphics[width=1\linewidth]{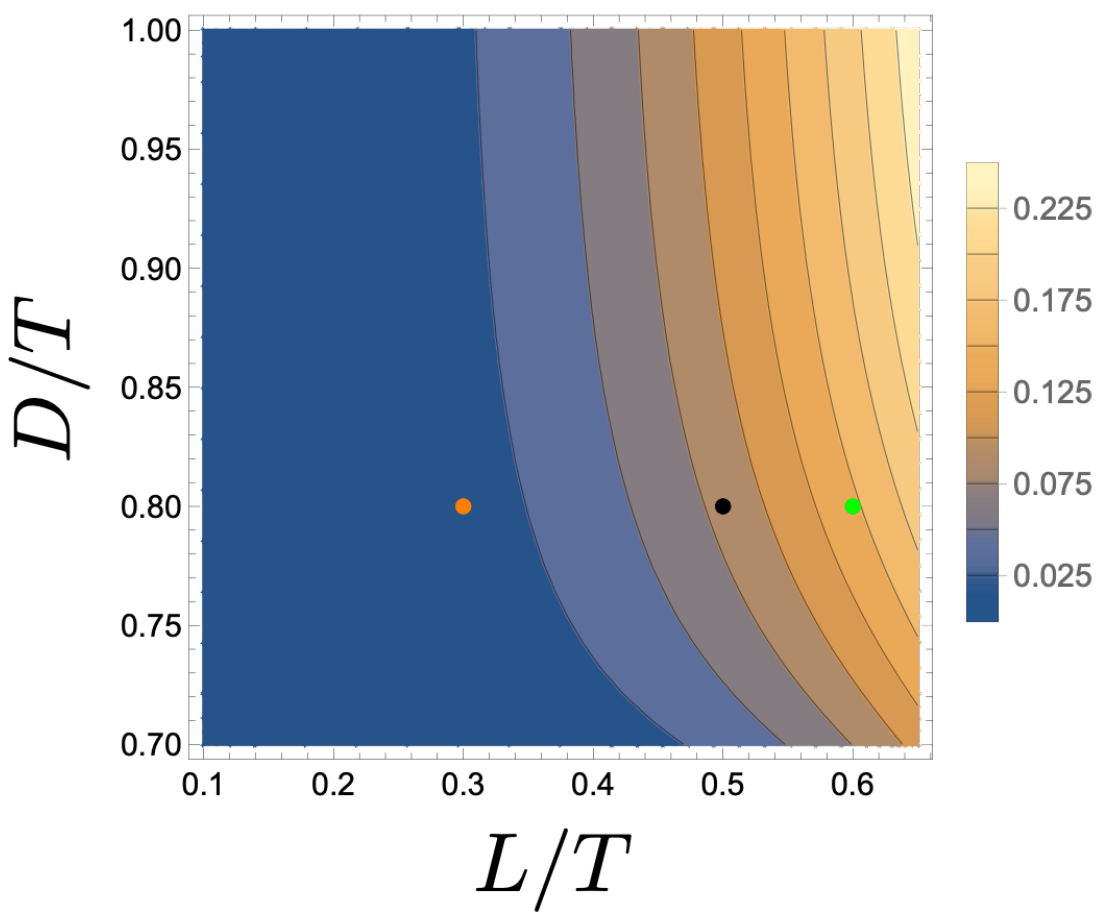}
  \end{minipage}
  \centering
  \begin{minipage}[b]{0.4\linewidth}
  \includegraphics[width=1\linewidth]{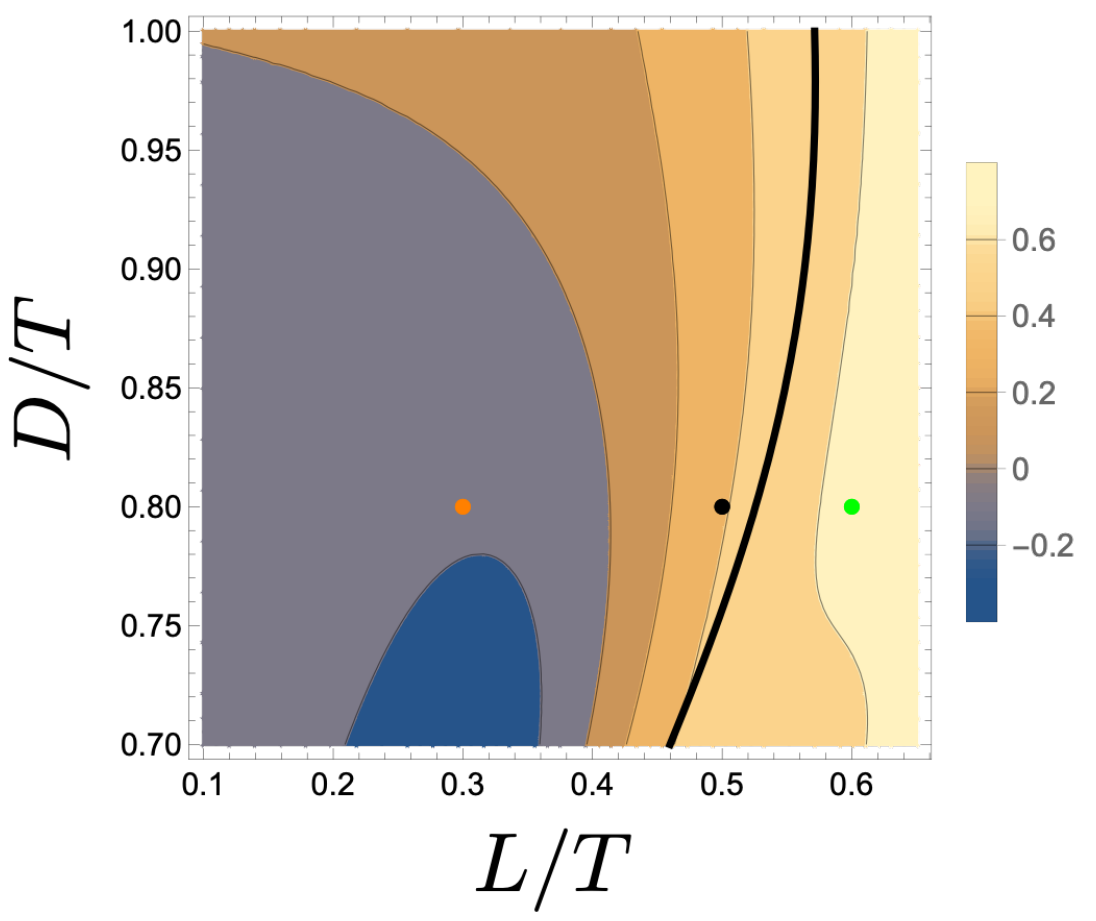}
  \end{minipage}
  \\
  \begin{minipage}[b]{0.32\linewidth}
  \includegraphics[width=1\linewidth]{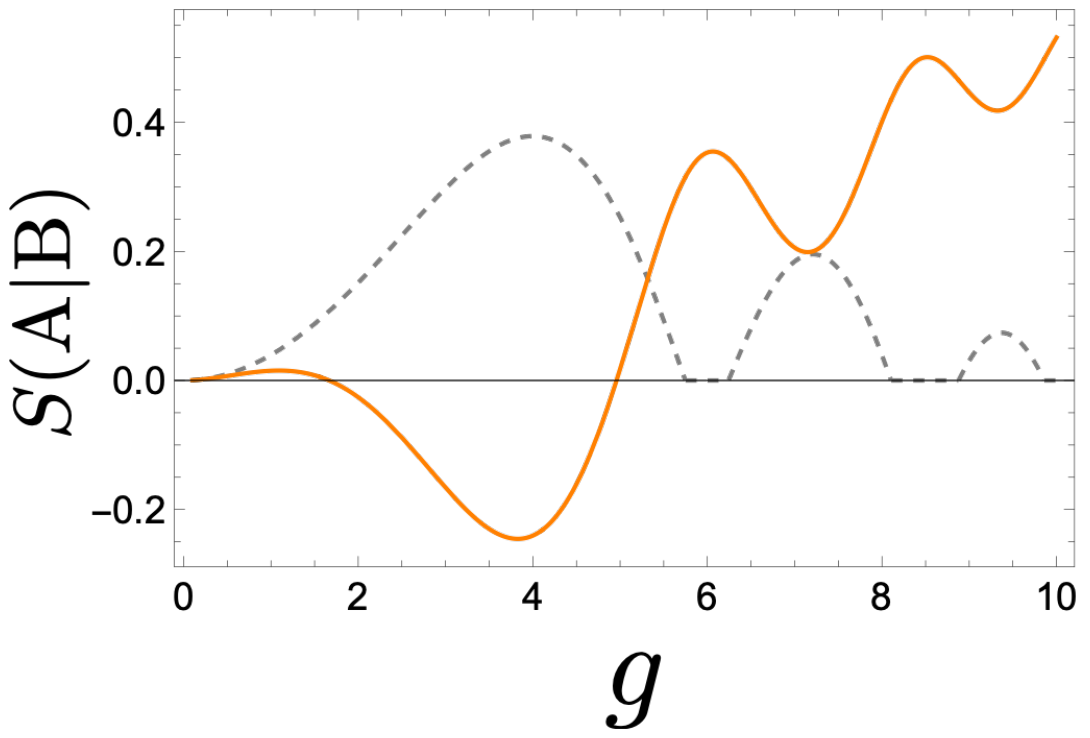}
  \end{minipage}
  \begin{minipage}[b]{0.32\linewidth}
  \includegraphics[width=1\linewidth]{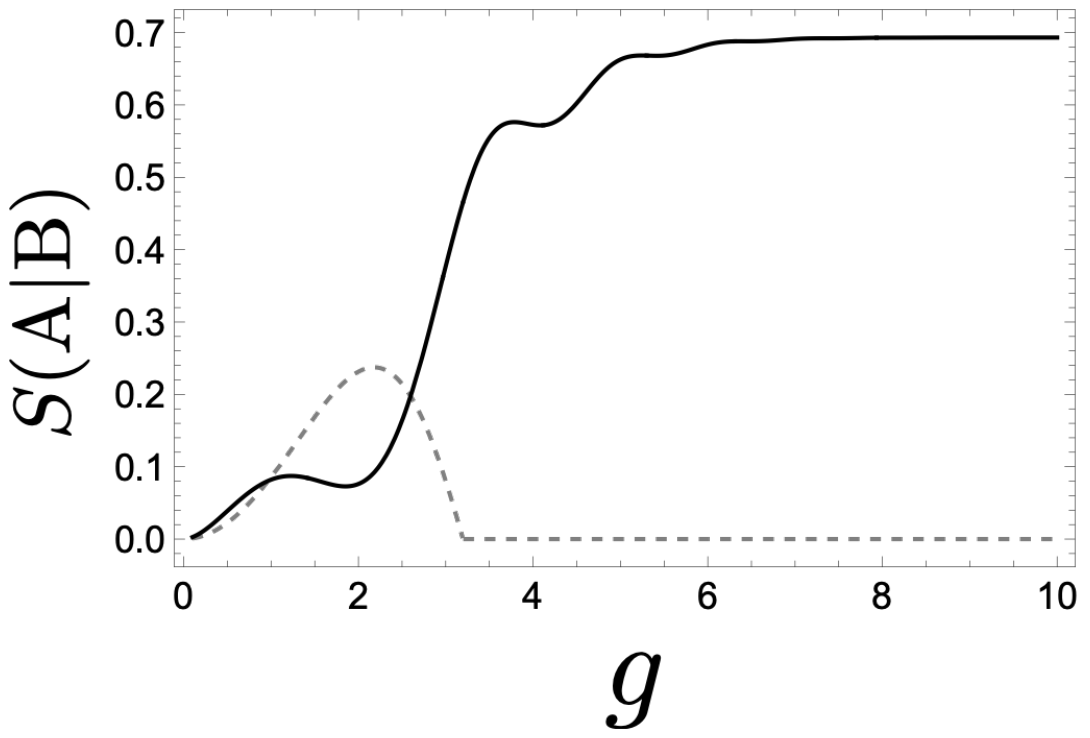}
  \end{minipage}
  \begin{minipage}[b]{0.32\linewidth}
  \includegraphics[width=1\linewidth]{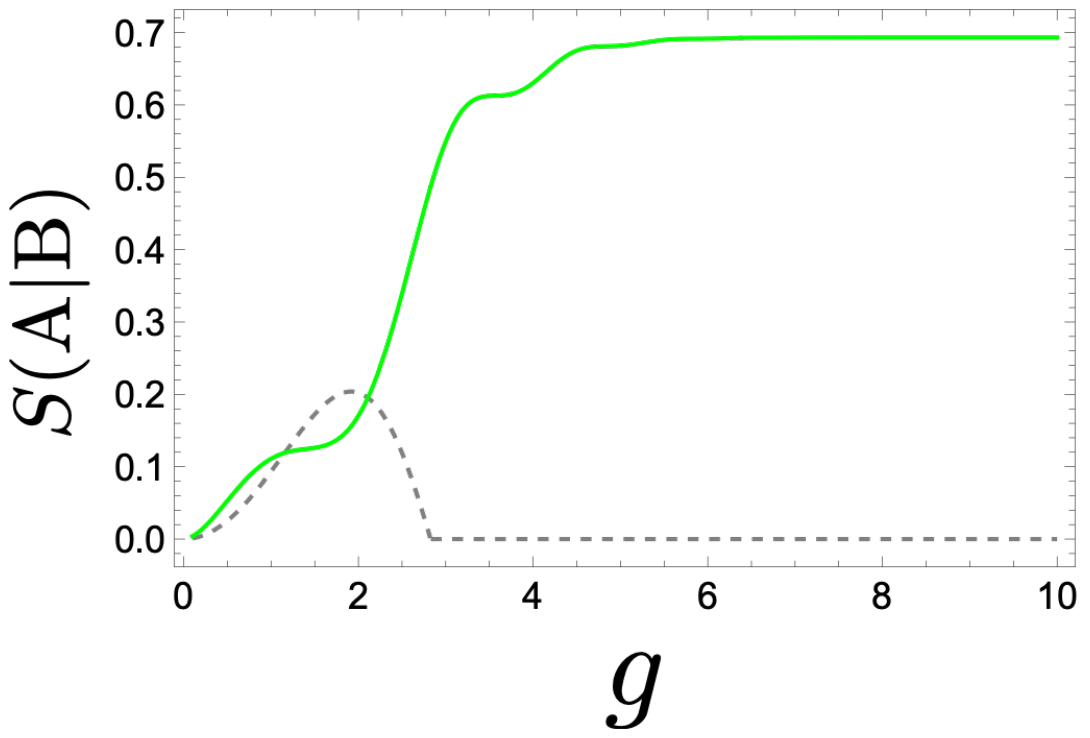}
  \end{minipage}
  \caption{
  Upper panel: Contour plot of the conditional von Neumann entropy $S(\text{A}|\text{B})$ as functions of $L/T$ (horizontal axis) and $D/T$ (vertical axis), where we adopted the coupling constant $g=1$ (left panel) and $g=3$ (right panel). 
  The orange, black, and green circle are points when $L/T=3/10$, $L/T=5/10$, and $L/T=6/10$, respectively, with $D/T=8/10$.
  The thick black curve in the right panel is the boundary where the negativity $\mathcal{N}_{\text{AB}}$ vanishes.
  Namely, the negativity $\mathcal{N}_{\text{AB}}$ vanishes in the right-hand region of the boundary. 
  Lower panel: The left, center, and right panels show the conditional von Neumann entropy $S(\text{A}|\text{B})$ as a function of the coupling constant $g$.
  The dashed black line in each panel depicts the negativity $\mathcal{N}_{\text{AB}}$.
  These graphs assume the parameters of the orange, black, and green circles, respectively, in the upper panel.
  \label{fig:TDLcondE}}
\end{figure}
The parameter dependence of the conditional von Neumann entropy is depicted in Fig.~\ref{fig:TDLcondE}.
The upper panels in Fig.~\ref{fig:TDLcondE} represent the contour plots of the conditional von Neumann entropy versus $L/T$ and $D/T$ with the coupling constant $g=1$ (left panel) and $g=3$ (right panel).
In the upper-right panel of Fig.~\ref{fig:TDLcondE}, the thick black curve shows the boundary of the entanglement generation between two particles, where the negativity $\mathcal{N}_{\text{AB}}$ vanishes
in the right region of the thick black curve.
In the left panel, 
$S(\text{A}|\text{B})>0$ is satisfied in the parameter region.
However, the right panel shows three regions:
$S(\text{A}|\text{B}) < 0$ and $\mathcal{N}_{\text{AB}}>0$, $S(\text{A}|\text{B}) > 0$ and $\mathcal{N}_{\text{AB}}>0$, and $S(\text{A}|\text{B}) > 0$ and $\mathcal{N}_{\text{AB}}=0$.
In the region $S(\text{A}|\text{B}) < 0$ and $\mathcal{N}_{\text{AB}}>0$, the conditional von Neumann entropy $S(\text{A}|\text{B})$ is negative;
therefore, we cannot judge whether particle A and gravitons are entangled or not from the inequality~\eqref{EoF} because the entanglement of formation $E_{\text{f}}(\rho_{\text{A,g}})$ may not be positive.
However, the negativity $\mathcal{N}_{\text{AB}}$ is positive, and then two particles A and B are entangled.
Regions $S(\text{A}|\text{B}) > 0$ and $\mathcal{N}_{\text{AB}}>0$ indicate that particle A and gravitons and two particles A and B are entangled.
In the regions $S(\text{A}|\text{B}) > 0$ and $\mathcal{N}_{\text{AB}}=0$, we can understand that two particles, A and B, are not entangled, but particle A and gravitons are in an entangled state.

The orange, black, and green marks in the upper panels of Fig.~\ref{fig:TDLcondE} represent the three typical classes, 
$S(\text{A}|\text{B}) < 0$ and $\mathcal{N}_{\text{AB}}>0$, $S(\text{A}|\text{B}) > 0$ and $\mathcal{N}_{\text{AB}}>0$, and $S(\text{A}|\text{B}) > 0$ and $\mathcal{N}_{\text{AB}}=0$, respectively. 
The lower three panels of Fig.~\ref{fig:TDLcondE} represent the conditional von Neumann entropy $S(\text{A}|\text{B})$ (solid curve) and the negativity $\mathcal{N}_{\text{AB}}$ (thin black dashed curve) as functions of $g$, for the three typical classes. 
It can be seen that the solid curve $S(\text{A}|\text{B})$ in each panel saturates at the coupling constant $g$, whereas the negativity vanishes due to the decoherence when $g$ becomes large.

\section{Behavior of Quantum discord}
Here, we investigate the behavior of the quantum superposition of the gravitational field using quantum discord~\cite{Ollivier, Xi, Luo}.
Quantum discord is a measure of all quantum correlations, including entanglement.
The quantum discord of the composite system AB is defined by the difference between the quantum mutual information 
$\mathcal{I}(\text{A}, \text{B})$ and the classical correlation $\mathcal{J}(\text{A}, \text{B})$
\begin{align}
\mathcal{D}(\text{A}, \text{B})
=
\mathcal{I}(\text{A}, \text{B})-\mathcal{J}(\text{A}, \text{B}).
\end{align}
The nonvanishing of the quantum discord is related to the quantum superposition principle~\cite{Ollivier}.
In particular, we focus on the quantum discord between particle A and the gravitational field 
$\mathcal{D}(\text{A, g})$, which may be the evidence of the quantum superposition of the gravitational field; that is, the quantumness of the gravitational field.
To simplify calculations, we represent $\mathcal{D}(\text{A, g})$ by using the entanglement of formation $E_{\text{f}}(\rho_{\text{AB}})$ and the conditional von Neumann entropy $S(\text{A}|\text{B})$ as
\begin{align}
\mathcal{D}(\text{A, g})
=
E_{\text{f}}(\rho_{\text{AB}})+S(\text{A}|\text{B}),
\label{discordin}
\end{align}
where the details of derivation are presented in Appendix~\ref{proofEoF}.
The above Eq.~\eqref{discordin} shows that the quantum correlation between the particle and the gravitational field is determined by the parameters of the systems A and B, which is one of the features of monogamy.
We introduce a formula of the entanglement of formation for a two-qubit system with respect to the two-qubit state $\rho_{\text{AB}}$ as \cite{Bennett,Wootters}
\begin{align}
E_{\text{f}}(\rho_{\text{AB}})
=
h
\left(
\frac{1+\sqrt{1-C^2(\rho_{\text{AB}})}}{2}
\right),
\end{align}
where we defined $h(x):=-x\log_{2} x-(1-x)\log_{2} (1-x)$, and
$C(\rho_{\text{AB}})$ is concurrence, which measures the degree of entanglement in the mixed state~\cite{Bennett, Hill, Wootters}.
The concurrence for the mixed state $\rho_{\text{AB}}$ of a qubit system is introduced as
\begin{align}
C(\rho_{\text{AB}}):=\text{max} \{0, \alpha_{1}-\alpha_{2}-\alpha_{3}-\alpha_{4}\}
\end{align}
with $\alpha_{1} \geq \alpha_{2} \geq \alpha_{3} \geq \alpha_{4}$.
Here $\alpha_{i}$ ($i=1, \ldots, 4$) are the square root of eigenvalues of the non-Hermitian matrix $\rho_{\text{AB}}(\sigma^{\text{A}}_{y}\otimes\sigma^{\text{B}}_{y})\rho^{*}_{\text{AB}}(\sigma^{\text{A}}_{y}\otimes\sigma^{\text{B}}_{y})$, where
$\rho^{*}_{\text{AB}}$ is the complex conjugate of $\rho_{\text{AB}}$, and $\sigma^{\text{A}}_{y}$ ($\sigma^{\text{B}}_{y}$) is the Pauli matrix, which works for the local system A (B).
In the following, we study the behavior of the quantum discord $\mathcal{D}(\text{A, g})$ in the two regions: $D \gg T \gg L$ and $T \gg D \gg L$.

\subsection{$D \gg T \gg L$ regime}
We first consider the case $D \gg T \gg L$.
In this regime, two particles A and B are not entangled, i.e., $E_{\text{f}}(\rho_{\text{AB}})=0$.
Thus, the quantum discord is exactly equivalent to the conditional von Neumann entropy $S(\text{A}|\text{B})$ based on Eq.~\eqref{discordin}.
\begin{figure}[t]
  \centering
  \begin{minipage}[b]{0.43\linewidth}
  \includegraphics[width=1\linewidth]{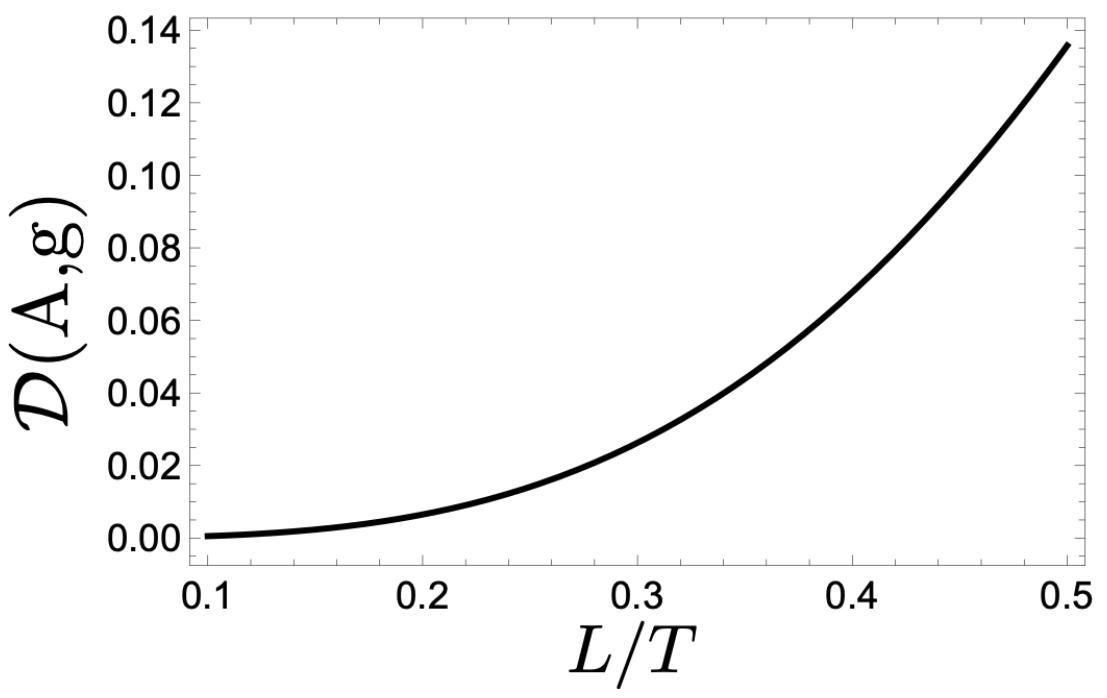}
  \end{minipage}\hspace{0.5cm}
  \begin{minipage}[b]{0.43\linewidth}
  \includegraphics[width=1\linewidth]{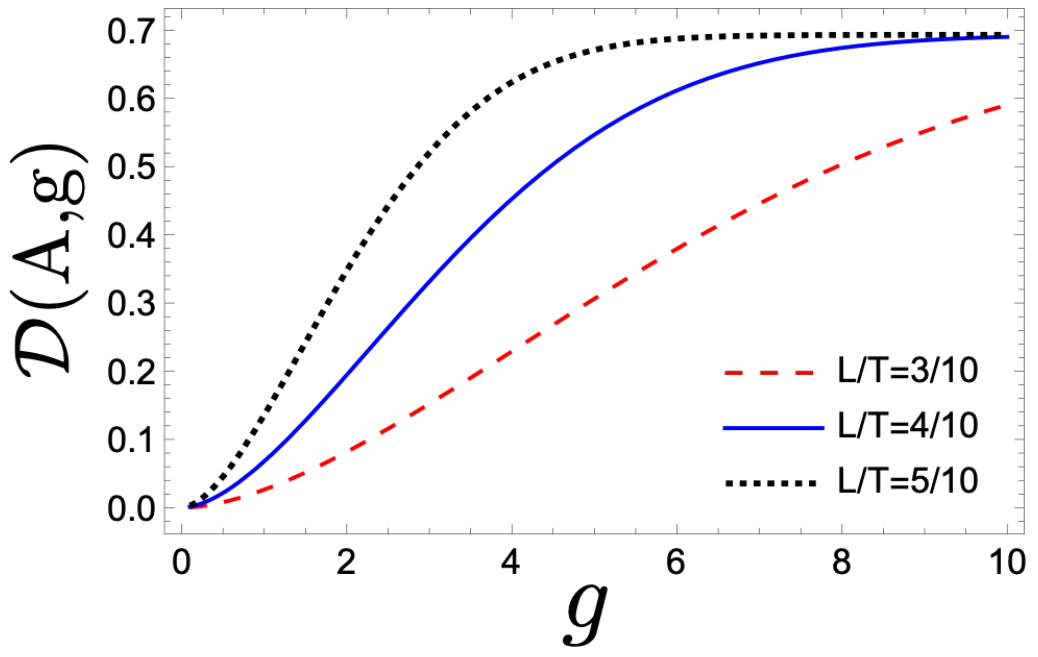}
  \end{minipage}
  \caption{
  Left panel represents the quantum discord $\mathcal{D}(\text{A}, \text{g})$ as a function of $L/T$ \cm{for} $g=1$. 
  Similarly, the right panel does the same as a function of the coupling $g$, 
  where we fixed $L/T=3/10$ (red dashed curve), $L/T=4/10$ (blue solid curve), and $L/T=5/10$ (black dotted curve). 
  In both panels, we adopted $D/T=5$.
  \label{fig:discorddtl}}
\end{figure}
Figure~\ref{fig:discorddtl} depicts the behavior of the quantum discord $\mathcal{D}(\text{A}, \text{g})$ as a function of $L/T$ (left panel) and the coupling constant $g$ (right panel).
The left panel shows that when the length scale of the superposition $L$ of particle A $L$ increases, the gravitational field also becomes well quantum superposition state.
The right panel of Fig.~\ref{fig:discorddtl} can be understood as follows.
As the coupling constant $g$ increases, the interaction between particle A and the gravitational field becomes stronger, and they become well-correlated. 
This results in the decoherence of particle A, and the entanglement between the two particles vanishes.
These relationships represent the monogamy property between the two particles and the gravitational field, as shown in Eq.~\eqref{result}.

\subsection{$T \gg D \gg L$ regime}

\begin{figure}[H]
  \centering
  \begin{minipage}[b]{0.43\linewidth}
  \includegraphics[width=1\linewidth]{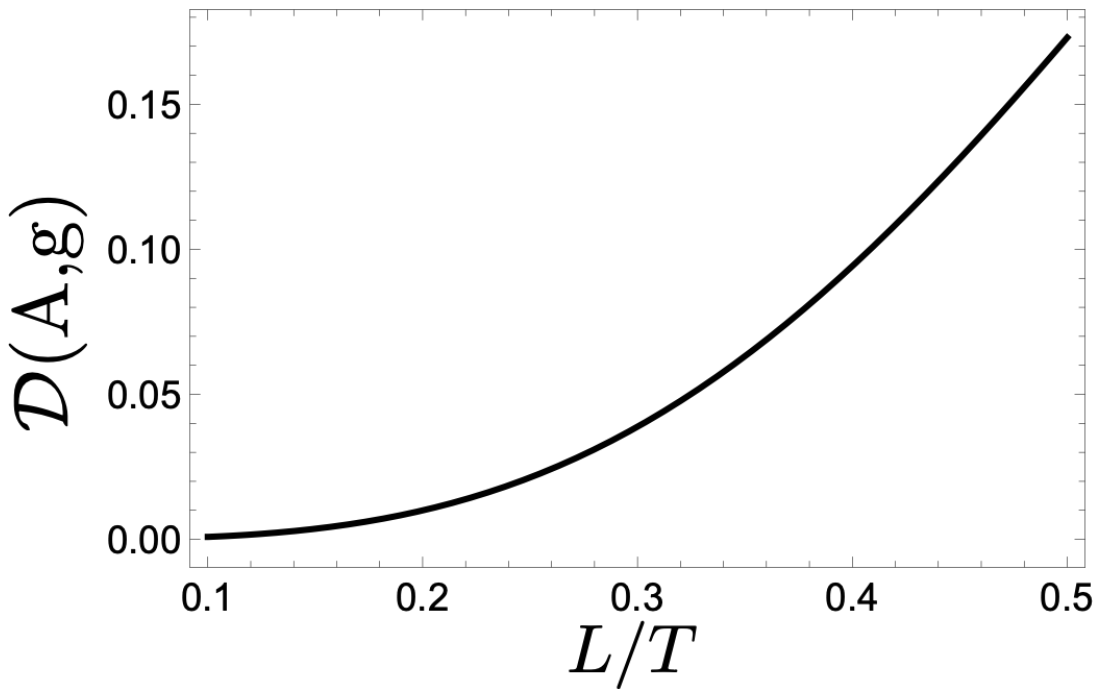}
  \end{minipage}\hspace{0.5cm}
  \begin{minipage}[b]{0.43\linewidth}
  \includegraphics[width=1\linewidth]{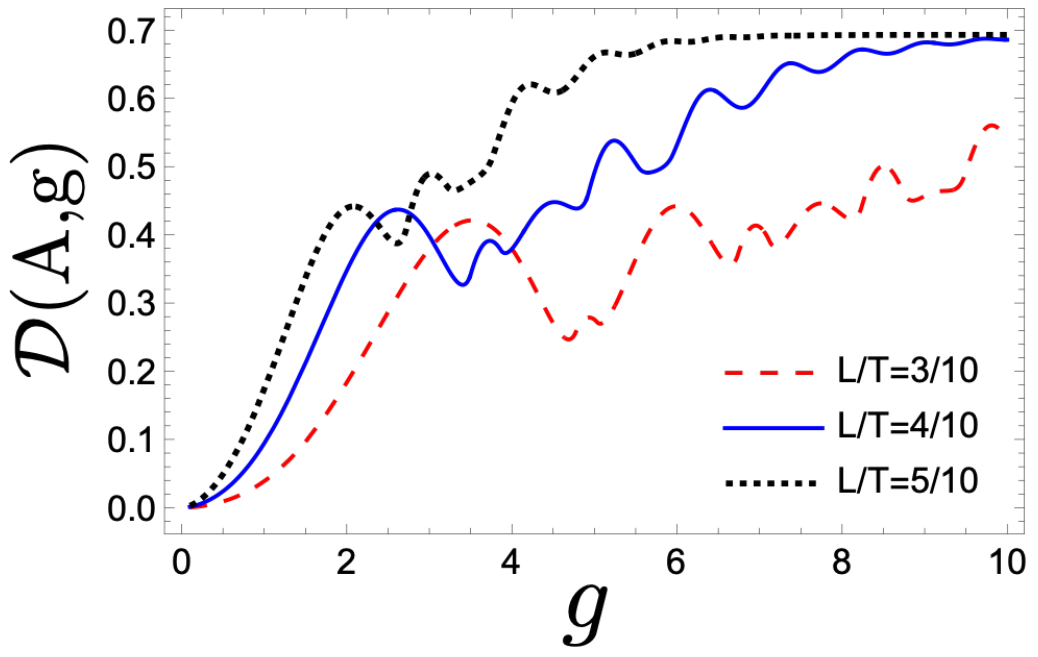}
  \end{minipage}
  \caption{
  The quantum discord $\mathcal{D}(\text{A}, \text{g})$ as a function of $L/T$ (left panel) with $g=1$ and $g$ (right panel) with $D/T=7/10$.
  \label{fig:discord}}
\end{figure}

Next, we consider the case $T \gg D \gg L$.
Fig.~\ref{fig:discord} is the same as Fig.~\ref{fig:discorddtl} but with
$D/T=7/10<1$, the quantum discord $\mathcal{D}(A, g)$ as a function of $L/T$ (left panel) and $g$ (right panel).
We come to the same conclusion in the region $D \gg T \gg L$ that increasing the superposition width of the particle A leads to the well-superposition state of the gravitational field.
Moreover, when the coupling constant $g$ increases, the decoherence becomes efficient in suppressing the entanglement generation between the two particles.
Note that, in this regime, the two particles A and B are slightly entangled, which reduces the correlation between particle A and the gravitational field.
From the viewpoint of monogamy, the suppression of the entanglement between two particles makes the entanglement between particle A and the gravitational field strong.

\section{conclusion}
Deepening our understanding of the quantumness of gravity will play a crucial role in the unification of the gravity theory and quantum mechanics.
To this end, this study focused on the quantum superposition of gravitational fields based on the quantum theory of linearized gravity.
We analyzed the dynamics of a two-particle system in each superposition state interacting with a gravitational field and revealed the entanglement structure between particle and the gravitational field.
We derived an inequality in which the conditional von Neumann entropy between two particles yields a lower bound on the entanglement between the particle and the gravitational field.
Furthermore, we found that the conditional von Neumann entropy has a trade-off relationship with the negativity between the two particles.
Thus, we showed that the particle and field are always entangled if the two particles are not entangled.
In addition, we evaluated quantum discord to quantitatively evaluate the quantum correlations between the particle and the gravitational field. 
Quantum discord characterizes the quantum superposition of the gravitational field. 
Consequently, the superposition of the gravitational field becomes more significant as the separation of the superposition states of the particles increases.


\acknowledgements
We thank Y. Nambu for the useful discussions.
Y.S. was supported by the Kyushu University Innovator Fellowship in Quantum Science.
A.M. was supported by JSPS KAKENHI (Grants No.~JP23K13103 and No.~JP23H01175).
K.Y. was supported by JSPS KAKENHI (Grants No.~JP22H05263 and No.~JP23H01175).

\begin{appendix}

\section{BRST FORMALISM IN QED}
Here, we briefly introduce the quantization of the QED using the BRST formalism~\cite{Sugiyama1} to preserve the covariance of the theory with a view toward its extension to quantum gravity theory.
The Lagrangian density of the QED in BRST formalism is written as follows:
\begin{equation}
\mathscr{L}=\mathscr{L}_{\text{QED}}+\mathscr{L}_{\text{GF}+\text{FP}},
\quad 
\mathscr{L}_{\text{QED}}=-\frac{1}{4} F_{\mu \nu} F^{\mu \nu}+\bar{\psi} (i\gamma^{\mu}D_{\mu}-m ) \psi,
\end{equation}
where 
$F_{\mu \nu}=\partial_{\mu} A_{\nu}-\partial_{\nu}A_{\mu}$ is the field strength of the $U(1)$ gauge field $A_{\mu}$, $\psi$ is the Dirac field with mass $m$, $\bar{\psi}=\psi^{\dagger}\gamma^{0}$, $\gamma^{\mu}$ is the gamma matrix satisfying $\{\gamma^{\mu}, \gamma^{\nu}\}=2\eta^{\mu \nu}$,  $D_{\mu}=\partial_{\mu}+ieA_{\mu}$ is the covariant derivative, which
includes the electromagnetic interaction term with the coupling constant $e$, and $\mathscr{L}_{{\text{GF}}+{\text{FP}}}$ is the gauge fixing and Faddeev-Popov ghost term.
The Lagrangian density $\mathscr{L}_{\text{QED}}$ is invariant under the following transformation
 \begin{equation}
 \psi \rightarrow e^{-ie\theta(x)}\psi \simeq (1-ie\theta(x))\psi \equiv \psi +\delta \psi, 
 \quad 
 A_{\mu} \rightarrow A_{\mu}+\partial_{\mu} \theta(x) \equiv A_{\mu}+\delta A_{\mu},
 \end{equation} 
where 
$\theta(x)$ is a real function and $\delta \psi$ and 
$\delta A_{\mu}$ are defined as 
$\delta \psi \equiv -ie\theta \psi$ 
and $\delta A_{\mu} \equiv \partial_{\mu} \theta$.
To give the gauge fixing and Faddeev-Popov ghost term $\mathscr{L}_{{\text{GF}}+{\text{FP}}}$
, we define $\theta(x)\equiv \lambda C(x)$, where $\lambda$ and $C(x)$ are the global and local Grassmann numbers. 
The field $C(x)$ is a scalar field, but it satisfies the anticommutation relations $\{C(x), C(y)\}=0$, which is the Faddeev-Popov ghost field.
We rewrite 
$\delta \psi$ and $\delta A_{\mu}$
as follows
\begin{align}
\delta \psi(x) =\lambda (-ieC(x)\psi(x)) \equiv \lambda \delta_{\text{B}}\psi(x),
\quad
\delta A_{\mu}=\lambda (\partial_{\mu}C(x))\equiv \lambda \delta_{\text{B}}A_{\mu},
\quad
\delta_{\text{B}}C(x)=0,
\label{BRST1}
\end{align}
where the operator $\delta_{\text{B}}$ is defined to satisfy the nilpotency $\delta^2_{\text{B}} =0$.
We also introduce the antighost field $\bar{C}(x)$ and the Nakanishi-Lautrup field $B(x)$, which satisfy
\begin{align}
\delta_{\text{B}} \bar{C}(x)=iB(x),
\quad
\delta_{\text{B}} B(x)=0.
\label{BRST2}
\end{align}
The transformation of Eqs.~\eqref{BRST1} and \eqref{BRST2} is referred to as the BRST transformation.
We can choose the gauge fixing and Faddeev-Popov ghost terms of the Lagrangian $\mathscr{L}_{{\text{GF}}+{\text{FP}}}$
as follows:
\begin{align}
\mathscr{L}_{\text{GF}+\text{FP}}=-i\delta_{\text{B}} (\bar{C}F),
\quad
F=\partial^{\mu} A_{\mu}+\frac{1}{2}\xi {\text{B}},
\end{align} 
where $\xi$ is an arbitrary parameter.
Thus, the full Lagrangian density in BRST formalism is
\begin{align}
\label{fulleom}
 \mathscr{L}&=-\frac{1}{4} F_{\mu \nu} F^{\mu \nu}+\bar{\psi} (i\gamma^{\mu}D_{\mu} \psi-m ) \psi+\frac{1}{2} \xi B^2 -\partial^{\mu}B A_{\mu}-i\partial^{\mu}\bar{C}\partial_{\mu}C.
\end{align}
The equations of motion for fields $A_{\mu}, B, C, \bar{C}$ are given by the Euler-Lagrange equations from the full Lagrangian,
\begin{align}
\label{aeom}
0&=\partial^{\nu}F_{\nu \mu}-J_{\mu}-\partial_{\mu}B,\\
\label{beom}
0&=\partial^{\mu}A_{\mu}+\xi B,\\
0&=\Box C= \Box \bar{C},
\label{ghost}
\end{align} 
where 
$J_{\mu}=e\bar{\psi}\gamma_{\mu}\psi$ is the matter current of the Dirac field.
The fields 
$C(x)$ and 
$\bar{C}(x)$ follow the free evolution so that they do not interact with the other fields.
Substituting \eqref{beom} into \eqref{fulleom}, we arrive at the following Lagrangian density,
\begin{align}
\mathscr{L}&=-\frac{1}{4} F_{\mu \nu} F^{\mu \nu}+\bar{\psi} (i\gamma^{\mu}D_{\mu} \psi-m ) \psi-\frac{1}{2\xi} (\partial_{\mu} A^{\mu})^2-i\partial^{\mu}\bar{C}\partial_{\mu}C,
\end{align}
and the BRST transformations are summarized as
\begin{align}
\delta_B A_{\mu}=\partial_{\mu} C,
\quad
\delta_{B}\psi =-ieC\psi, \delta_B C=0, \delta_B \bar{C} =\frac{i}{\xi}(\partial_{\mu}A^{\mu}).
\end{align}
Because of the BRST transformation, the Lagrangian density has a global symmetry (BRST symmetry)
\begin{align}
\lambda \delta_{B}\mathscr{L}=0.
\end{align}
Thanks to this global symmetry, there is a conserved current referred to as the BRST current ${J}^{\mu}_\text{B}$ defined by
\begin{align}
J^{\mu}_{\text{B}}=\sum_{\text{I}} \frac{\partial \mathscr{L}}{\partial(\partial_{\mu} \Phi_{\text{I}})} \delta_{\text{B}}\Phi_{\text{I}}
=-F^{\mu \nu}\partial_\nu C-\frac{1}{\xi} \partial_\nu A^\nu \partial^\mu C+J^\mu C
,
\end{align}
where 
$\Phi_\text{I}=\{A_\mu, \psi, C, \bar{C}\}$.
The BRST charge 
$Q_\text{B}$ is given by
\begin{align}
Q_\text{B} \equiv \int d^3x J^0_\text{B}(x)=\int d^3x \Big[(\partial_i C)F^{i0}+J^0 C-\frac{1}{\xi} (\partial_{\mu}A ^{\mu})\dot{C}
\Big].
\end{align}
We perform the canonical quantization procedure in the Feynman gauge ($\xi =1$). 
The canonical conjugate momenta are defined as 
\begin{align}
\pi^{\mu}_{A}\equiv \frac{\partial \mathscr{L}}{\partial \dot{A}_{\mu}}
=-F^{0\mu}-(\partial_{\nu}A^{\nu})\eta^{0\mu}
\ 
,
\pi_{\psi}\equiv \frac{\partial \mathscr{L}}{\partial \dot{\psi}}
=i\bar{\psi}\gamma^{0}
\ 
,
\pi_{c}\equiv \frac{\partial \mathscr{L}}{\partial \dot{C}}
=i\dot{\bar{C}}
\ ,
\pi_{\bar{c}}\equiv \frac{\partial \mathscr{L}}{\partial \dot{\bar{C}}}
=i\dot{C},
\end{align}
where $``\cdot"$ denotes the derivative with respect to time $x^{0}=t$.
The equal-time commutation relations are assigned as follows
\begin{align}
\{\hat{\psi}(x), \hat{\pi}_{\psi}(y)\}|_{x^0=y^0}=i\delta^3(\mathbf{x}-\mathbf{y}),
\nonumber\\
\{\hat{C}(x), \hat{\pi}_{c}(y)\}|_{x^0=y^0}=i\delta^3(\mathbf{x}-\mathbf{y}),
\nonumber\\
\{\hat{\bar{C}}(x), \hat{\pi}_{\bar{c}}(y)\}|_{x^0=y^0}=i\delta^3(\mathbf{x}-\mathbf{y}),
\nonumber\\
[\hat{A}_{\mu}(x), \hat{\pi}^{\nu}_{A}(y)]|_{x^0=y^0}=i\delta^{\nu}_{\mu}\delta^3(\mathbf{x}-\mathbf{y})
.\nonumber
\end{align}
The quantized BRST charge is then given by 
\begin{equation}
\hat{Q}_\text{B}=\int d^3x[(\partial_i \hat{C})\hat{F}^{i0}+\hat{J}^0\hat{C}-(\partial_{\mu}\hat{A}^{\mu})\dot{\hat{C}}]=\int d^3x[-(\partial_{i}\hat{\pi}^{i})\hat{C}+\hat{J}^{0}\hat{C}+i\hat{\pi}^{0}\hat{\pi}_{\bar{c}}].
\label{intscr}
\end{equation}

It is well-known that when we quantize a gauge theory while maintaining the Lorentz covariance, a state space $\mathcal{V}$ with an indefinite metric is required.
For the standard probabilistic interpretation of quantum mechanics , a physical state 
$|\Psi_\text{phys} \rangle$ 
has no negative norm. 
This state with a non-negative norm is identified by imposing the following condition (the BRST condition)
\begin{align}
\hat{Q}_\text{B}|\Psi_{\text{phys}}\rangle=0,
\label{eq:QPsi=0}
\end{align}
where the physical state $|\Psi_{\text{phys}}\rangle$ satisfies $\langle \Psi_{\text{phys}}|\Psi_{\text{phys}}\rangle \geq0$.

\subsection{BRST charge in the interaction picture and in the Schr\"odinger picture}
We derive a useful form of the BRST charge for our computation. 
Using Eq.~\eqref{intscr}, we obtain the BRST charge in the interaction picture,
\begin{align}
\hat{Q}^{\text{I}}_\text{B}(t)
=
e^{i\hat{H}_0t}\hat{Q}_\text{B} e^{-i\hat{H}_0t}
=
\int d^3x
[-(\partial_{i}\hat{\pi}^{i\text{I}})\hat{C}^{\text{I}}
+\hat{J}^{0}_{\text{I}}\hat{C}^{\text{I}}+i\hat{\pi}^{0\text{I}}\hat{\pi}^{\text{I}}_{\bar{c}}],
\label{intqb}
\end{align}
where 
$\hat{\phi}^{\text{I}}
=
e^{i\hat{H}_0t}\hat{\phi} e^{-i\hat{H}_0t}$, 
$\hat{\phi}
=
\{\hat{A}_{\mu}, \hat{\pi}^{\mu}, \hat{C}, \hat{\bar{C}}, \hat{\pi}_{c}, \hat{\pi}_{\bar{c}}, \hat{J}^{0}\}$, 
and they satisfy the following Heisenberg equation
\begin{align}
i\dot{\hat{\phi}}^{\text{I}}=[\hat{\phi}^{\text{I}}, \hat{H}_{0}].
\end{align}
Here $\hat{H}_{0}$ is the free Hamiltonian derived from the Lagrangian~\eqref{fulleom}.
The gauge field $\hat{A}^{\text{I}}_{\mu}(x)$ and the ghost field $\hat{C}^{\text{I}}(x)$ satisfy the Klein-Gordon equation.
The solutions are the superposition of the usual plane-wave solutions as follows:
\begin{align}
\label{solutiona}
\hat{A}^{\text{I}}_{\mu}(x)&=\int \frac{d^3k}{\sqrt{(2\pi)^32k^{0}}}(\hat{a}_{\mu}(\bm{k})e^{ik\cdot x}+\text{H.c.}),\\
\label{solutionc}
\hat{C}^{\text{I}}(x)&=\int \frac{d^3k}{\sqrt{(2\pi)^32k^{0}}}(\hat{c}(\bm{k})e^{ik\cdot x}+\text{H.c.}),
\end{align}
where 
$k^0=|\bm{k}|$, $\hat{a}_{\mu}(\bm{k})$ 
and
$\hat{c}(\bm{k})$ are the annihilation operators of the gauge field 
$\hat{A}^{\text{I}}_{\mu}(x)$, 
and the ghost field $\hat{C}^{\text{I}}(x)$, respectively. 
The annihilation operators 
$\hat{a}_{\mu}(\bm{k})$, $\hat{c}(\bm{k})$, 
and the creation operators satisfy the following commutation relations
\begin{equation}
\bigl[\hat{a}_{\mu}(\bm{k}),\hat{a}^{\dagger}_{\nu}(\bm{k}^{\prime})\bigr]=\eta_{\mu \nu}\delta (\bm{k}-\bm{k}^{\prime}), 
\quad
\big\{\hat{c}(\bm{k}),\hat{c}^{\dagger}(\bm{k}^{\prime})\bigr\}=\delta (\bm{k}-\bm{k}^{\prime}).
\end{equation}
Substituting \eqref{solutiona} and \eqref{solutionc} into Eq.~\eqref{intqb}, we obtain the BRST charge in the interaction picture
\begin{align}
\hat{Q}^{\text{I}}_\text{B}(t)
=
\int \frac{d^3k}{\sqrt{(2\pi)^3}} \left[\left(k^{\mu}\hat{a}_{\mu}(\bm{k})+\frac{\hat{\tilde{J}}^{0}_{\text{I}}(t,\bm{k})}{\sqrt{2k^0}}e^{ik^0t}\right)c^{\dagger}(\bm{k})+\text{H.c.}\right],
\label{intqb2}
\end{align}
where $\hat{\tilde{J}}^{0}_{\text{I}}(t,\bm{k})$ is the Fourier transformation of $\hat{J}^{0}_{\text{I}}(t,\bm{x})$
\begin{align}
\hat{J}^{0}_{\text{I}}(t,\bm{x})=\int \frac{d^3k}{\sqrt{(2\pi)^3}}\hat{\tilde{J}}^{0}_{\text{I}}(t,\bm{k})e^{i\bm{k}\cdot \bm{x}}.
\end{align}
Using the BRST charge in the interaction picture~\eqref{intqb} and \eqref{intqb2}, the BRST charge in the Schr\"{o}dinger picture is obtained as
\begin{align}
\hat{Q}_\text{B}=e^{-i\hat{H}_0t}\hat{Q}^{\text{I}}_\text{B}(t)e^{i\hat{H}_0t}=\int \frac{d^3k}{\sqrt{(2\pi)^3}} \left[\left(k^{\mu}\hat{a}_{\mu}(\bm{k})+\frac{\hat{\tilde{J}}^{0}(\bm{k})}{\sqrt{2k^0}}\right)c^{\dagger}(\bm{k})+\text{H.c.}\right],
\label{eq:QBSch}
\end{align}
where we used
\begin{equation}
e^{-i\hat{H}_0t}\hat{a}_{\mu}(\bm{k})e^{i\hat{H}_0t}=\hat{a}_{\mu}(\bm{k})e^{ik^0t}, 
\quad 
e^{-i\hat{H}_0t}\hat{c}^{\dagger}(\bm{k})e^{i\hat{H}_0t}=\hat{c}^{\dagger}(\bm{k})e^{-ik^0t},
\quad 
e^{-i\hat{H}_0t}\hat{\tilde{J}}^0_{\text{I}}(t, \bm{k})e^{i\hat{H}_0t}=\hat{\tilde{J}}^0(\bm{k}),
\end{equation}
Here, 
$\hat{\tilde{J}}^0(\bm{k})$ 
denotes the Fourier transform of the matter current in the Schr\"{o}dinger picture.

\subsection{BRST condition for two charged particles}

Using the explicit form of the BRST charge in the Schr\"{o}dinger picture~\eqref{eq:QBSch}, we derive the BRST conditions for our model~\eqref{hamiltonian}, which consists of two charged particles interacting with an electromagnetic field.
Assuming a physical state $|\Psi_{\text{phys}}\rangle=|\Psi^{\prime}_{\text{phys}}\rangle \otimes |0\rangle_{c}$, where $|0\rangle_{c}$ is the ground state of the ghost field, and using \eqref{eq:QBSch}, we can reduce the BRST condition \eqref{eq:QPsi=0} as 
\begin{equation}
\label{coh}
\Big(k^{\mu}\hat{a}_{\mu}(\bm{k})+\frac{\hat{\tilde{J}}^{0}(\bm{k})}{\sqrt{2k^{0}}}
\Big)
|\Psi^{\prime}_{\text{phys}} \rangle 
= 0
\end{equation}
with the current operator of the two charged particles A and B
$\hat{\tilde{J}}^{0}(\bm{k})
=\hat{\tilde{J}}^{0}_{\text{A}}(\bm{k})
+\hat{\tilde{J}}^{0}_{\text{B}}(\bm{k})$ in Fourier space.
When 
$|\Psi^{\prime}_{\text{phys}}\rangle$
is the initial state given in~\eqref{inistate1}, \eqref{coh} leads to the following equation,
\begin{align}
0&=\Big(k^{\mu}\hat{a}_{\mu}(\bm{k})+\frac{\hat{\tilde{J}}^{0}(\bm{k})}{\sqrt{2k^{0}}}
\Big)|\Psi^{\prime}_{\text{phys}} \rangle 
\nonumber 
\\
&= \Big(
k^{\mu}\hat{a}_{\mu}(\bm{k})+\frac{\hat{\tilde{J}}^{0}(\bm{k})}{\sqrt{2k^{0}}}
\Big)
\frac{1}{{2}}
\big(|\text{R}\rangle_{\text{A}}
+ |\text{L}\rangle_{\text{A}}
\big) 
\otimes
\big(|\text{R}\rangle_{\text{B}}
+ |\text{L}\rangle_{\text{B}}
\big) 
\otimes |\alpha \rangle_\text{EM}
\nonumber 
\\
&
\approx \frac{1}{{2}}
\big(|\text{R}\rangle_{\text{A}}
+ |\text{L}\rangle_{\text{A}}
\big) 
\otimes
\big(|\text{R}\rangle_{\text{B}}
+ |\text{L}\rangle_{\text{B}}
\big) 
\otimes  
\Big(k^{\mu}\hat{a}_{\mu}(\bm{k})+\frac{\tilde{J}^{0}(\bm{k})}{\sqrt{2k^{0}}}
\Big) |\alpha \rangle_\text{EM},
\end{align}
where the approximation \eqref{approx3} was used in the last line, and 
$\tilde{J}^{0}(\bm{k})
=\tilde{J}^0_\text{A}(\bm{k})+\tilde{J}^0_\text{B}(\bm{k})$ 
with
$\tilde{J}^0_\text{A}(\bm{k})
=\tilde{J}^0_\text{AR}(\bm{k})+\tilde{J}^0_\text{AL}(\bm{k})
$
and
$\tilde{J}^0_\text{B}(\bm{k})
=\tilde{J}^0_\text{BR}(\bm{k})+\tilde{J}^0_\text{BL}(\bm{k})$ 
are the currents in Fourier space in the trajectories R and L of charged particles A and B, respectively.
Here the state $|\alpha \rangle_\text{EM}$ is 
$|\alpha \rangle_\text{EM}=D(\alpha)|0\rangle_{\text{EM}}$. 
Hence the initial coherent state of the electromagnetic field must satisfy
\begin{align}
\label{coh2}
\Big(k^{\mu}\hat{a}_{\mu}(\bm{k})+\frac{\tilde{J}^{0}(\bm{k})}{\sqrt{2k^{0}}}
\Big) |\alpha\rangle_\text{EM}=0. 
\end{align}
Because the displacement operator $\hat{D}(\alpha)$ given in \eqref{D} has the following relation
\begin{align}
\hat{D}^{\dagger}(\alpha)\hat{a}_{\mu}(\bm{k})\hat{D}(\alpha)=\hat{a}_{\mu}(\bm{k})+\alpha_{\mu}(\bm{k}),
\label{DdgaD}
\end{align}
we obtain the constraint for the complex function 
$\alpha^\mu(\bm{k})$ as 
\begin{equation}
k^{\mu} \alpha_{\mu}(\bm{k})=-\frac{\tilde{J}^{0}(\bm{k})}{\sqrt{2k^{0}}}, 
\end{equation}
where we used $\hat{a}_{\mu}(\bm{k})|0\rangle_{\text{EM}}=0$.
This is the BRST condition for the model of two charged particles presented in the main text.
Thus, to ensure the BRST condition, the complex function $\alpha_{\mu}(\bm{k})$ must satisfy the condition~\eqref{brstcondition}.

\section{SUMMARY OF QED CASE FORMULAS AND UNIFIED DESCRIPTION WITH GRAVITATIONAL FIELD}
Here we present the eigenvalues of the density matrix $\rho^{\text{EM}}_{\text{A}}$, $\rho^{\text{EM}}_{\text{B}}$, and $\rho^{\text{EM}}_{\text{AB}}$ to derive the von Neumann entropy.
Further, we also give the formula of the minimum eigenvalue of the partial transposed density matrix of $\rho^{\text{EM}}_{\text{AB}}$.
Then, to compute the quantum discord in Sec. V, we introduce the result of the concurrence.
Finally, these quantities are summarized via the unified notation of QED and the gravitational field versions.
\subsection{Results of the eigenvalues of the density matrix 
$\rho^{\text{EM}}_{\text{A}}$, $\rho^{\text{EM}}_{\text{B}}$, and $\rho^{\text{EM}}_{\text{AB}}$}
The eigenvalues of the density matrix 
$\rho^{\text{EM}}_{\text{A}}=\text{Tr}_{\text{B}}[\rho^{\text{EM}}_{\text{AB}}]$, 
$\rho^{\text{EM}}_{\text{B}}=\text{Tr}_{\text{A}}[\rho^{\text{EM}}_{\text{AB}}]$ 
and $\rho^{\text{EM}}_{\text{AB}}$ are directly obtained as
\begin{align}
\lambda_{\pm}[\rho^{\text{EM}}_{\text{A}}]
&=
\frac{1}{2}
\left[
1\pm e^{-\Gamma^{\text{EM}}_{\text{A}}}\cos
\Big[
\frac{\Phi^{\text{EM}}_{\text{AB}}}{2}
\Big]
\right],
\quad
\lambda_{\pm}[\rho^{\text{EM}}_{\text{B}}]
=
\frac{1}{2}
\left[
1\pm e^{-\Gamma^{\text{EM}}_{\text{B}}}\cos
\Big[
\frac{\Phi^{\text{EM}}_{\text{BA}}}{2}
\Big]
\right],
\label{valueAB}
\\
\quad
\lambda_{\pm}[\rho^{\text{EM}}_{\text{AB}}]
&=
\frac{1}{4}
\Big[
1-e^{-\Gamma^{\text{EM}}_{\text{A}}-\Gamma^{\text{EM}}_{\text{B}}} \cosh[\Gamma^{\text{EM}}_\text{c}]
\pm
\Big\{\big(e^{-\Gamma^{\text{EM}}_{\text{A}}}-e^{-\Gamma^{\text{EM}}_{\text{B}}}\big)^2
\nonumber\\
\quad
&
+4e^{-\Gamma^{\text{EM}}_{\text{A}}-\Gamma^{\text{EM}}_{\text{B}}}
\sin^2
\Big[
\frac{\Phi^{\text{EM}}_{\text{AB}}-\Phi^{\text{EM}}_{\text{BA}}}{4}
\Big]
+e^{-2\Gamma^{\text{EM}}_{\text{A}}-2\Gamma^{\text{EM}}_{\text{B}}} \sinh^2[\Gamma^{\text{EM}}_\text{c}]\Big\}^{\frac{1}{2}}
\Big],
\label{lambdam}
\\
\quad
\lambda'_{\pm}[\rho^{\text{EM}}_{\text{AB}}]
&=
\frac{1}{4}
\Big[
1+e^{-\Gamma^{\text{EM}}_{\text{A}}-\Gamma^{\text{EM}}_{\text{B}}} \cosh[\Gamma^{\text{EM}}_\text{c}]
\pm 
\Big\{
\big(e^{-\Gamma^{\text{EM}}_{\text{A}}}-e^{-\Gamma^{\text{EM}}_{\text{B}}}\big)^2
\nonumber\\
\quad
&
+4e^{-\Gamma^{\text{EM}}_{\text{A}}-\Gamma^{\text{EM}}_{\text{B}}}
\cos^2
\Big[
\frac{\Phi^{\text{EM}}_{\text{AB}}-\Phi^{\text{EM}}_{\text{BA}}}{4}
\Big]
+e^{-2\Gamma^{\text{EM}}_{\text{A}}-2\Gamma^{\text{EM}}_{\text{B}}} \sinh^2[\Gamma^{\text{EM}}_\text{c}]
\Big\}^{\frac{1}{2}}
\Big],
\label{lambdap}
\end{align}
where 
$\Gamma^\text{EM}_{i} \, (i=\text{A},\text{B})$
are introduced in Eqs.~\eqref{gammaiEM}, and the quantities $\Phi^{\text{EM}}_{\text{AB}}$ and $\Phi^{\text{EM}}_{\text{BA}}$ are defined as
\begin{align}
\Phi^{\text{EM}}_{\text{AB}}
&=
\int d^4x d^4y
\Delta J^{\mu}_{\text{A}}(x)\Delta J^{\nu}_{\text{B}}(y)
G^{\text{r}}_{\mu\nu}(x,y),
\quad
\Phi^{\text{EM}}_{\text{BA}}
=
\int d^4x d^4y
\Delta J^{\mu}_{\text{B}}(x)\Delta J^{\nu}_{\text{A}}(y)
G^{\text{r}}_{\mu\nu}(x,y).
\label{phiabEM}
\end{align}
The quantities $\Gamma^{\text{EM}}_{\text{c}}$ 
is represented as
\begin{align}
\Gamma^{\text{EM}}_\text{c}
&=
\frac{1}{2}
\int d^4x \int d^4y 
\Delta J^{\mu}_{\text{A}}(x) \Delta J^{\nu}_{\text{B}}(y)
\langle \{\hat{A}^\text{I}_{\mu}(x), \hat{A}^\text{I}_{\nu}(y)\}\rangle.
\label{gammacEM}
\end{align}

\subsection{Result of the minimum eigenvalue of the density matrix $(\rho^{\text{EM}}_{\text{AB}})^{\text{T}_{\text{A}}}$}
Here we present the formula of the minimum eigenvalue of the density matrix $(\rho^{\text{EM}}_{\text{AB}})^{\text{T}_{\text{A}}}$.
The eigenvalues of the density matrix 
$(\rho^{\text{EM}}_{\text{AB}})^{\text{T}_{\text{A}}}$ 
are 
\begin{align}
\lambda_{\pm}
[(\rho^{\text{EM}}_{\text{AB}})^{\text{T}_{\text{A}}}]
&=
\frac{1}{4}
\Big[
1-e^{-\Gamma^{\text{EM}}_{\text{A}}-\Gamma^{\text{EM}}_{\text{B}}} \cosh[\Gamma^{\text{EM}}_\text{c}]
\pm
\Big\{
\big(e^{-\Gamma^{\text{EM}}_{\text{A}}}-e^{-\Gamma^{\text{EM}}_{\text{B}}}\big)^2
\nonumber\\
\quad
&
+4e^{-\Gamma^{\text{EM}}_{\text{A}}-\Gamma^{\text{EM}}_{\text{B}}}
\sin^2\Big[
\frac{\Phi^{\text{EM}}_{\text{AB}}+\Phi^{\text{EM}}_{\text{BA}}}{4}
\Big]
+e^{-2\Gamma^{\text{EM}}_{\text{A}}-2\Gamma^{\text{EM}}_{\text{B}}} \sinh^2[\Gamma^{\text{EM}}_\text{c}]
\Big\}^{\frac{1}{2}}
\Big],
\\
\quad
\lambda'_{\pm}
[(\rho^{\text{EM}}_{\text{AB}})^{\text{T}_{\text{A}}}]
&=
\frac{1}{4}
\Big[
1+e^{-\Gamma^{\text{EM}}_{\text{A}}-\Gamma^{\text{EM}}_{\text{B}}} \cosh[\Gamma^{\text{EM}}_\text{c}]
\pm
\Big\{
\big(e^{-\Gamma^{\text{EM}}_{\text{A}}}-e^{-\Gamma^{\text{EM}}_{\text{B}}}\big)^2
\nonumber\\
\quad
&
+4e^{-\Gamma^{\text{EM}}_{\text{A}}-\Gamma^{\text{EM}}_{\text{B}}}
\cos^2
\Big[
\frac{\Phi^{\text{EM}}_{\text{AB}}+\Phi^{\text{EM}}_{\text{BA}}}{4}
\Big]
+e^{-2\Gamma^{\text{EM}}_{\text{A}}-2\Gamma^{\text{EM}}_{\text{B}}} \sinh^2[\Gamma^{\text{EM}}_\text{c}]
\Big\}^{\frac{1}{2}}
\Big].
\end{align}
Note that 
$\lambda_{-}[(\rho^{\text{EM}}_{\text{AB}})^{\text{T}_{\text{A}}}]$
is the minimum eigenvalue 
$\lambda^{\text{EM}}_{\text{min}}$
\begin{align}
\lambda^{\text{EM}}_{\text{min}}
&=
\frac{1}{4}
\Big[
1-e^{-\Gamma^{\text{EM}}_{\text{A}}-\Gamma^{\text{EM}}_{\text{B}}} \cosh[\Gamma^{\text{EM}}_\text{c}]
-\Big\{
\big(e^{-\Gamma^{\text{EM}}_{\text{A}}}-e^{-\Gamma^{\text{EM}}_{\text{B}}}\big)^2
\nonumber\\
\quad
&
+4e^{-\Gamma^{\text{EM}}_{\text{A}}-\Gamma^{\text{EM}}_{\text{B}}}
\sin^2
\Big[
\frac{\Phi^{\text{EM}}_{\text{AB}}+\Phi^{\text{EM}}_{\text{BA}}}{4}
\Big]
+e^{-2\Gamma^{\text{EM}}_{\text{A}}-2\Gamma^{\text{EM}}_{\text{B}}} \sinh^2[\Gamma^{\text{EM}}_\text{c}]
\Big\}^{\frac{1}{2}}
\Big].
\label{negativityEM}
\end{align}

\subsection{Result of the concurrence}
We present the results of the concurrence.
As mentioned in the main text, the concurrence between two particles A and B is
\begin{align}
C(\rho^{\text{EM}}_{\text{AB}}):=\text{max} \{0, \alpha^{\text{EM}}_{1}-\alpha^{\text{EM}}_{2}-\alpha^{\text{EM}}_{3}-\alpha^{\text{EM}}_{4}\}
\end{align}
with 
$\alpha^{\text{EM}}_{1} \geq \alpha^{\text{EM}}_{2} \geq \alpha^{\text{EM}}_{3} \geq \alpha^{\text{EM}}_{4}$.
Here 
$\alpha^{\text{EM}}_{i}$ ($i=1,\ldots ,4$)
are the square root of eigenvalues of the non-Hermitian density matrix 
$\rho^{\text{EM}}_{\text{AB}}(\sigma^{\text{A}}_{y}\otimes\sigma^{\text{B}}_{y})
(\rho^{\text{EM}}_{\text{AB}})^{*}(\sigma^{\text{A}}_{y}\otimes\sigma^{\text{B}}_{y})$,
where $(\rho^{\text{EM}}_{\text{AB}})^{*}$ 
is the complex conjugate of 
$\rho^{\text{EM}}_{\text{AB}}$, and 
$\sigma^{\text{A}}_{y}$ ($\sigma^{\text{B}}_{y}$) is the Pauli matrix for local system A (B).
The eigenvalues of 
$\alpha^{\text{EM}}_{i}$ ($i=1, \dots ,4$)
are given as follows:
\begin{align}
(\alpha^{\text{EM}}_{1})^2
=
\frac{1}{16}
&
\Big[
1+2e^{-\Gamma^{\text{EM}}_{\text{A}}-
\Gamma^{\text{EM}}_{\text{B}}}
\cosh[\Gamma^{\text{EM}}_{\text{c}}]
+e^{-2\Gamma^{\text{EM}}_{\text{A}}-2\Gamma^{\text{EM}}_{\text{B}}}
\cosh[2\Gamma^{\text{EM}}_{\text{c}}]
-e^{-2\Gamma^{\text{EM}}_{\text{A}}}
\cos[\Phi^{\text{EM}}_{\text{AB}}]
-e^{-2\Gamma^{\text{EM}}_{\text{B}}}
\cos[\Phi^{\text{EM}}_{\text{BA}}]
\nonumber\\
\quad
&
-2e^{-\Gamma^{\text{EM}}_{\text{A}}-\Gamma^{\text{EM}}_{\text{B}}}
\cos
\Big[
\frac{\Phi^{\text{EM}}_{\text{AB}}+\Phi^{\text{EM}}_{\text{BA}}}{4}
\Big]
+
\Big\{
-4e^{-2\Gamma^{\text{EM}}_{\text{A}}-2\Gamma^{\text{EM}}_{\text{B}}}
(\cosh[\Gamma^{\text{EM}}_{\text{c}}]
-\cos
\Big[
\frac{\Phi^{\text{EM}}_{\text{AB}}-\Phi^{\text{EM}}_{\text{BA}}}{4}
\Big]
\nonumber\\
\quad
&
+2\sinh[\Gamma^{\text{EM}}_{\text{A}}]
\sinh[\Gamma^{\text{EM}}_{\text{B}}])^2
+
\Big(
1+2e^{-\Gamma^{\text{EM}}_{\text{A}}-\Gamma^{\text{EM}}_{\text{B}}}
\cosh[\Gamma^{\text{EM}}_{\text{c}}]
+e^{-2\Gamma^{\text{EM}}_{\text{A}}-2\Gamma^{\text{EM}}_{\text{B}}}
\cosh[2\Gamma^{\text{EM}}_{\text{c}}]
\nonumber\\
\quad
&
-e^{-2\Gamma^{\text{EM}}_{\text{A}}}
\cos[\Phi^{\text{EM}}_{\text{AB}}]
-e^{-2\Gamma^{\text{EM}}_{\text{B}}}
\cos[\Phi^{\text{EM}}_{\text{BA}}]
-2e^{-\Gamma^{\text{EM}}_{\text{A}}-\Gamma^{\text{EM}}_{\text{B}}}
\cos
\Big[
\frac{\Phi^{\text{EM}}_{\text{AB}}+\Phi^{\text{EM}}_{\text{BA}}}{4}
\Big]
\Big)^2
\Big\}^{\frac{1}{2}}
\Big],
\label{alpha1}
\end{align}
\begin{align}
(\alpha^{\text{EM}}_{2})^2
=
\frac{1}{16}
&
\Big[
1+2e^{-\Gamma^{\text{EM}}_{\text{A}}-
\Gamma^{\text{EM}}_{\text{B}}}
\cosh[\Gamma^{\text{EM}}_{\text{c}}]
+e^{-2\Gamma^{\text{EM}}_{\text{A}}-2\Gamma^{\text{EM}}_{\text{B}}}
\cosh[2\Gamma^{\text{EM}}_{\text{c}}]
-e^{-2\Gamma^{\text{EM}}_{\text{A}}}
\cos[\Phi^{\text{EM}}_{\text{AB}}]
-e^{-2\Gamma^{\text{EM}}_{\text{B}}}
\cos[\Phi^{\text{EM}}_{\text{BA}}]
\nonumber\\
\quad
&
-2e^{-\Gamma^{\text{EM}}_{\text{A}}-\Gamma^{\text{EM}}_{\text{B}}}
\cos
\Big[
\frac{\Phi^{\text{EM}}_{\text{AB}}+\Phi^{\text{EM}}_{\text{BA}}}{4}
\Big]
-
\Big\{
-4e^{-2\Gamma^{\text{EM}}_{\text{A}}-2\Gamma^{\text{EM}}_{\text{B}}}
(\cosh[\Gamma^{\text{EM}}_{\text{c}}]
-\cos
\Big[
\frac{\Phi^{\text{EM}}_{\text{AB}}-\Phi^{\text{EM}}_{\text{BA}}}{4}
\Big]
\nonumber\\
\quad
&
+2\sinh[\Gamma^{\text{EM}}_{\text{A}}]
\sinh[\Gamma^{\text{EM}}_{\text{B}}])^2
+
\Big(
1+2e^{-\Gamma^{\text{EM}}_{\text{A}}-\Gamma^{\text{EM}}_{\text{B}}}
\cosh[\Gamma^{\text{EM}}_{\text{c}}]
+e^{-2\Gamma^{\text{EM}}_{\text{A}}-2\Gamma^{\text{EM}}_{\text{B}}}
\cosh[2\Gamma^{\text{EM}}_{\text{c}}]
\nonumber\\
\quad
&
-e^{-2\Gamma^{\text{EM}}_{\text{A}}}
\cos[\Phi^{\text{EM}}_{\text{AB}}]
-e^{-2\Gamma^{\text{EM}}_{\text{B}}}
\cos[\Phi^{\text{EM}}_{\text{BA}}]
-2e^{-\Gamma^{\text{EM}}_{\text{A}}-\Gamma^{\text{EM}}_{\text{B}}}
\cos
\Big[
\frac{\Phi^{\text{EM}}_{\text{AB}}+\Phi^{\text{EM}}_{\text{BA}}}{4}
\Big]
\Big)^2
\Big\}^{\frac{1}{2}}
\Big],
\label{alpha2}
\end{align}
\begin{align}
(\alpha^{\text{EM}}_{3})^2
=
\frac{1}{16}
&
\Big[
1-2e^{-\Gamma^{\text{EM}}_{\text{A}}-
\Gamma^{\text{EM}}_{\text{B}}}
\cosh[\Gamma^{\text{EM}}_{\text{c}}]
+e^{-2\Gamma^{\text{EM}}_{\text{A}}-2\Gamma^{\text{EM}}_{\text{B}}}
\cosh[2\Gamma^{\text{EM}}_{\text{c}}]
-e^{-2\Gamma^{\text{EM}}_{\text{A}}}
\cos[\Phi^{\text{EM}}_{\text{AB}}]
-e^{-2\Gamma^{\text{EM}}_{\text{B}}}
\cos[\Phi^{\text{EM}}_{\text{BA}}]
\nonumber\\
\quad
&
+2e^{-\Gamma^{\text{EM}}_{\text{A}}-\Gamma^{\text{EM}}_{\text{B}}}
\cos
\Big[
\frac{\Phi^{\text{EM}}_{\text{AB}}+\Phi^{\text{EM}}_{\text{BA}}}{4}
\Big]
+
\Big\{
-4e^{-2\Gamma^{\text{EM}}_{\text{A}}-2\Gamma^{\text{EM}}_{\text{B}}}
(\cosh[\Gamma^{\text{EM}}_{\text{c}}]
-\cos
\Big[
\frac{\Phi^{\text{EM}}_{\text{AB}}-\Phi^{\text{EM}}_{\text{BA}}}{4}
\Big]
\nonumber\\
\quad
&
-2\sinh[\Gamma^{\text{EM}}_{\text{A}}]
\sinh[\Gamma^{\text{EM}}_{\text{B}}])^2
+
\Big(
1+2e^{-\Gamma^{\text{EM}}_{\text{A}}-\Gamma^{\text{EM}}_{\text{B}}}
\cosh[\Gamma^{\text{EM}}_{\text{c}}]
+e^{-2\Gamma^{\text{EM}}_{\text{A}}-2\Gamma^{\text{EM}}_{\text{B}}}
\cosh[2\Gamma^{\text{EM}}_{\text{c}}]
\nonumber\\
\quad
&
-e^{-2\Gamma^{\text{EM}}_{\text{A}}}
\cos[\Phi^{\text{EM}}_{\text{AB}}]
-e^{-2\Gamma^{\text{EM}}_{\text{B}}}
\cos[\Phi^{\text{EM}}_{\text{BA}}]
-2e^{-\Gamma^{\text{EM}}_{\text{A}}-\Gamma^{\text{EM}}_{\text{B}}}
\cos
\Big[
\frac{\Phi^{\text{EM}}_{\text{AB}}+\Phi^{\text{EM}}_{\text{BA}}}{4}
\Big]
\Big)^2
\Big\}^{\frac{1}{2}}
\Big],
\label{alpha3}
\end{align}
\begin{align}
(\alpha^{\text{EM}}_{4})^2
=
\frac{1}{16}
&
\Big[
1-2e^{-\Gamma^{\text{EM}}_{\text{A}}-
\Gamma^{\text{EM}}_{\text{B}}}
\cosh[\Gamma^{\text{EM}}_{\text{c}}]
+e^{-2\Gamma^{\text{EM}}_{\text{A}}-2\Gamma^{\text{EM}}_{\text{B}}}
\cosh[2\Gamma^{\text{EM}}_{\text{c}}]
-e^{-2\Gamma^{\text{EM}}_{\text{A}}}
\cos[\Phi^{\text{EM}}_{\text{AB}}]
-e^{-2\Gamma^{\text{EM}}_{\text{B}}}
\cos[\Phi^{\text{EM}}_{\text{BA}}]
\nonumber\\
\quad
&
+2e^{-\Gamma^{\text{EM}}_{\text{A}}-\Gamma^{\text{EM}}_{\text{B}}}
\cos
\Big[
\frac{\Phi^{\text{EM}}_{\text{AB}}+\Phi^{\text{EM}}_{\text{BA}}}{4}
\Big]
-
\Big\{
-4e^{-2\Gamma^{\text{EM}}_{\text{A}}-2\Gamma^{\text{EM}}_{\text{B}}}
(\cosh[\Gamma^{\text{EM}}_{\text{c}}]
-\cos
\Big[
\frac{\Phi^{\text{EM}}_{\text{AB}}-\Phi^{\text{EM}}_{\text{BA}}}{4}
\Big]
\nonumber\\
\quad
&
-2\sinh[\Gamma^{\text{EM}}_{\text{A}}]
\sinh[\Gamma^{\text{EM}}_{\text{B}}])^2
+
\Big(
1+2e^{-\Gamma^{\text{EM}}_{\text{A}}-\Gamma^{\text{EM}}_{\text{B}}}
\cosh[\Gamma^{\text{EM}}_{\text{c}}]
+e^{-2\Gamma^{\text{EM}}_{\text{A}}-2\Gamma^{\text{EM}}_{\text{B}}}
\cosh[2\Gamma^{\text{EM}}_{\text{c}}]
\nonumber\\
\quad
&
-e^{-2\Gamma^{\text{EM}}_{\text{A}}}
\cos[\Phi^{\text{EM}}_{\text{AB}}]
-e^{-2\Gamma^{\text{EM}}_{\text{B}}}
\cos[\Phi^{\text{EM}}_{\text{BA}}]
-2e^{-\Gamma^{\text{EM}}_{\text{A}}-\Gamma^{\text{EM}}_{\text{B}}}
\cos
\Big[
\frac{\Phi^{\text{EM}}_{\text{AB}}+\Phi^{\text{EM}}_{\text{BA}}}{4}
\Big]
\Big)^2
\Big\}^{\frac{1}{2}}
\Big].
\label{alpha4}
\end{align}

\subsection{Unified description of QED and gravitational field}
As shown in Eqs.~\eqref{valueAB}, \eqref{lambdam}, \eqref{lambdap}, \eqref{negativityEM}, \eqref{alpha1}, \eqref{alpha2}, \eqref{alpha3}, and \eqref{alpha4}, these quantities can be described using the quantities 
$\Gamma^{\text{EM}}_{\text{i}} \, (i=\text{A},\text{B})$, 
$\Gamma^{\text{EM}}_{\text{c}}$, 
$\Phi^{\text{EM}}_{\text{AB}}$, and 
$\Phi^{\text{EM}}_{\text{BA}}$.
By replacing these quantities with 
$\Gamma^\text{GR}_{i} \, (i=\text{A},\text{B})$, 
$\Gamma^\text{GR}_{\text{c}}$, 
$\Phi^\text{GR}_{\text{AB}}$ and $\Phi^\text{GR}_{\text{BA}}$, we obtain the formulas for gravitational case.
The quantities 
$\Gamma^\text{GR}_{i} \, (i=\text{A},\text{B})$, 
$\Gamma^\text{GR}_{\text{c}}$
are
\begin{align}
\Gamma^{\text{GR}}_{i}
&=
\frac{1}{4}\int d^4x\int d^4y 
\Delta T^{\mu\nu}_{i}(x)\Delta T^{\rho\sigma}_{i}(y)
\langle 
\{\hat{h}^\text{I}_{\mu\nu}(x), \hat{h}^\text{I}_{\rho\sigma}(y)\}
\rangle,
\\
\quad
\Gamma^{\text{GR}}_\text{c}
&=
\frac{1}{2}\int d^4x\int d^4y
\Delta T^{\mu\nu}_{\text{A}}(x)\Delta T^{\rho\sigma}_{\text{B}}(y)
\langle 
\{\hat{h}^\text{I}_{\mu\nu}(x), \hat{h}^\text{I}_{\rho\sigma}(y)\}
\rangle,
\end{align}
where we defined 
$\Delta T^{\mu\nu}_{i}=T^{\mu\nu}_{i\text{R}}-T^{\mu\nu}_{i\text{L}}$
with $i=\text{A}, \text{B}$.
The gravitational case of the quantities 
$\Phi^{\text{GR}}_{\text{AB}}$ and $\Phi^{\text{GR}}_{\text{BA}}$ 
can be expressed as
\begin{align}
\Phi^{\text{GR}}_{\text{AB}}
&=
\int d^4x d^4y
\Delta T^{\mu\nu}_{\text{A}}(x)\Delta T^{\rho\sigma}_{\text{B}}(y)
G^{\text{r}}_{\mu\nu\rho\sigma}(x,y),
\quad
\Phi^{\text{GR}}_{\text{BA}}
=
\int d^4x d^4y
\Delta T^{\mu\nu}_{\text{B}}(x)\Delta T^{\rho\sigma}_{\text{A}}(y)
G^{\text{r}}_{\mu\nu\rho\sigma}(x,y).
\end{align}
Subsequently, we adopt simple notations
$\Gamma_i\, (i=\text{A},\text{B})$, 
$\Gamma_\text{c}$, 
$\Phi_\text{AB}$ 
and 
$\Phi_\text{BA}$ to describe the quantities above for the electromagnetic and gravitational cases in a unified manner.

\section{CALCULATION OF $\Gamma^{\text{EM}}_{\text{A}}$, $\Gamma^{\text{EM}}_{\text{B}}$, AND $\Gamma^{\text{EM}}_{\text{c}}$\label{QED}}
Here, we exemplify the results of our calculations in 
$\Gamma^{\text{EM}}_{\text{A}}$, $\Gamma^{\text{EM}}_{\text{B}}$, and $\Gamma^{\text{EM}}_{\text{c}}$
to understand the physical meaning of these quantities.
Note that we recover the constants $c$ and $\hbar$ when we show the result of the calculation to emphasize that it is a dynamical field effect.
We first calculate the quantities 
$\Gamma^{\text{EM}}_{\text{A}}$ and $\Gamma^{\text{EM}}_{\text{B}}$.
We assume the following trajectories for particles A and B
\begin{equation}
X^\mu_\text{P} (t)  
=
[t,\epsilon_{\text{P}}X(t), 0,0 \Big]^\text{T},
\quad 
\epsilon_{\text{R}}
=
-\epsilon_{\text{L}}=1, 
\quad 
X(t)
=
8L\Big(1-\frac{t}{T} \Big)^2 \Big(\frac{t}{T}\Big)^2.
\label{xt}
\end{equation}
Because of the time and spatial translation invariance of the vacuum state, $\Gamma^{\text{EM}}_i$ 
is independent of the choice of the origin.
Thus, we can evaluate 
$\Gamma^{\text{EM}}_{\text{A}}$ and $\Gamma^{\text{EM}}_{\text{B}}$
by using the formula of Eq.~\eqref{gammaiEM} as
\begin{align}
\Gamma^{\text{EM}}_{\text{A}}
=
\Gamma^{\text{EM}}_{\text{B}}
&=
\frac{e^2}{4}
\oint_\text{C} dx^\mu \oint_\text{C} dy^\mu 
\langle \bigl\{
\hat{A}^\text{I}_\mu (x), \hat{A}^\text{I}_\nu (y) 
\bigr\}\rangle 
\nonumber \\
&\approx
\frac{e^2}{4}
\oint_\text{C} dx^\mu \oint_\text{C} dy^\mu 
\langle \bigl\{
\hat{A}^\text{I}_\mu (x^0, \bm{0}), \hat{A}^\text{I}_\nu (y^0, \bm{0}) 
\bigr\}\rangle 
\nonumber \\
&=
\frac{e^2}{4}
\oint_\text{C} dx^\mu \oint_\text{C} dy^\mu 
\frac{\eta_{\mu \nu} }{4\pi^2} 
\Big(
\frac{1}{-(t-t'-i\epsilon)^2}+\frac{1}{-(t-t'+i\epsilon)^2} 
\Big) 
\nonumber \\
&=
\frac{e^2}{16\pi^2}
\int^T_0 dt 
\Big(
\frac{d X^\mu_\text{R}}{dt}-\frac{d X^\mu_\text{L}}{dt}
\Big)
\int^T_0 dt'
\Big(
\frac{d X_{\text{R}\,\mu}}{dt'}-\frac{d X_{\text{L}\,\mu}}{dt'}
\Big) 
\Big(
\frac{1}{-(t-t'-i\epsilon)^2}+\frac{1}{-(t-t'+i\epsilon)^2} 
\Big) \nonumber \\
&=
\frac{e^2}{16 \pi^2}
\int^T_0 dt \int^T_0 dt'
\Big(
\frac{d \bm{X}_\text{R}}{dt}-\frac{d \bm{X}_\text{L}}{dt}
\Big) 
\cdot
\Big(
\frac{d \bm{X}_\text{R}}{dt'}-\frac{d \bm{X}_\text{L}}{dt'}
\Big) 
\Big(
\frac{1}{-(t-t'-i\epsilon)^2}+\frac{1}{-(t-t'+i\epsilon)^2} 
\Big)
\nonumber\\
&=
\frac{32e^2}{3\pi^2\hbar c} 
\Big(
\frac{L}{cT}
\Big)^2,
\label{gammaiem}
\end{align}
where we took the limit 
$\epsilon \rightarrow 0$ 
after the integration, and in the second line we used the dipole approximation~\cite{Mazzitelli, Hsiang} which ignores the spatial dependence of the photon field.
The dipole approximation is valid when the wavelength of the photon field $\lambda_{\text{p}}=T$ is considerably larger than the typical size ($\sim L$) of the region  where the charge exists.
This condition is always satisfied if we assume a nonrelativistic velocity $L/T \ll 1$.
Note that the quantities $\Gamma_{\text{A}}$ and $\Gamma_{\text{B}}$ vanish when we take the nonrelativistic limit $c\to \infty$.
Thus they represent relativistic corrections originating from the dynamical component of the electromagnetic field.

This decoherence can be interpreted in the following two ways.
The first is the emission of photons, which is estimated by using the Larmor formula for the power of radiation emitted from a nonrelativistic charged particle during the time $T$. 
The second is the vacuum fluctuation due to the photon field, which induces the dephasing of the charged particle.
The above two interpretations give the equivalent result discussed in Ref.~\cite{Sugiyama1}.

In the following, we calculate the quantity $\Gamma^{\text{EM}}_{\text{c}}$ by using Eq.~\eqref{gammacEM}, which characterizes the correlation between the two particles.
This quantity depends on the configuration of two particles, so we divide two regimes $D \gg T \gg L$ and $T \gg D \gg L$ and show the result of $\Gamma^{\text{EM}}_{\text{c}}$.

\subsubsection{\label{cphilTDL}$T \gg D \gg L$ regime}
Then, we focus on the regime
$T \gg D \gg L$. 
In this regime, two particles are causally connected and can communicate with each other.
The trajectories of two particles A and B are supposed to be
\begin{equation}
X^\mu_{\text{AP}}
=
[t,\epsilon_{\text{P}}X(t), 0,0]^\text{T}, 
\quad 
X^\mu_{\text{BQ}}(t)
=
[t,\epsilon_{\text{Q}}X(t)+D,0,0 \Big]^\text{T},
\quad 
\epsilon_{\text{R}}
=
-\epsilon_{\text{L}}=1, 
\quad 
X(t)
=
8L
\Big(
1-\frac{t}{T} 
\Big)^2 
\Big(
\frac{t}{T}
\Big)^2.
\end{equation}
In this configuration, we obtain the result of the 
$\Gamma^{\text{EM}}_{\text{c}}$ as
\begin{align}
\Gamma^{\text{EM}}_{c}
&=
\frac{e^2}{2
}\oint_{\text{C}_\text{A}} dx^\mu \oint_{\text{C}_\text{B}} dy^\nu  
\, 
\langle\{
\hat{A}^\text{I}_{\mu}(x), \hat{A}^\text{I}_{\nu}(y)
\}\rangle
\approx 
\frac{64e^2}{3\pi^2\hbar c} 
\Big(
\frac{L}{T}
\Big)^2
\Big\{
1+4\Big(\frac{D}{c T}\Big)^2 \ln \Big[\frac{D}{c T}\Big] 
\Big\}
\approx
\frac{64e^2}{3\pi^2\hbar c} 
\Big(
\frac{L}{c T}
\Big)^2
,
\label{gammactdl}
\end{align}
where, in the last approximation, we neglected the term proportional to $D/T$ because of the condition $T \gg D \gg L$.
The detailed derivation is also presented in~\cite{Sugiyama1}.
The results of the electromagnetic case in Eqs.~\eqref{eq:GammacA1} and~\eqref{gammactdl} are extended to the result of the gravitational case~\eqref{dtl} and~\eqref{tdl}.

\subsubsection{\label{cphilDTL}$D \gg T \gg L$ regime}
We first focus on the regime $D \gg T \gg L$ and calculate the quantity $\Gamma_{\text{c}}$.
We assume the following trajectories for the two charged particles, A and B:
\begin{equation}
X^\mu_{\text{AP}}(t)
=
\Big[t,\epsilon_{\text{P}}X(t), 0,0 \Big]^{T}, 
\quad 
X^\mu_{\text{BQ}}(t)
=
\Big[t,\epsilon_{\text{Q}}X(t-D)+D, 0,0 \Big]^{T}, 
\quad 
\epsilon_{\text{R}}=-\epsilon_{\text{L}}=1, 
\quad 
X(t)
=
8L\Big(1-\frac{t}{T} \Big)^2 \Big(\frac{t}{T}\Big)^2,
\end{equation}
where 
$X^\mu_{\text{BQ}}$ is defined in 
$D \leq t \leq T+D$.
In this configuration, the two particles are causally connected.
However, the particle B does not affect the system of particle A system. 
The quantity $\Gamma^{\text{EM}}_{\text{c}}$ is computed as
\begin{align}
\Gamma^{\text{EM}}_{c}
=\frac{e^2}{2}
\oint_{\text{C}_\text{A}} dx^\mu \oint_{\text{C}_\text{B}} dy^\nu  
\, 
\langle \{
\hat{A}^\text{I}_{\mu}(x), \hat{A}^\text{I}_{\nu}(y)
\}\rangle
\approx
-\frac{32e^2}{225\pi^2\hbar c} 
\Big(\frac{L}{c T}\Big)^2
\Big(\frac{c T}{D}\Big)^4.
\label{eq:GammacA1}
\end{align}
The detailed calculation is presented in~\cite{Sugiyama1}.

\section{DERIVATION OF THE TWO-POINT FUNCTION OF THE GRAVITATIONAL FIELD}
Here we derive the two-point function of the gravitational field.
The analysis in QED~\cite{Sugiyama1} suggested that only the dynamical components of the electromagnetic field contributes to the quantities $\Gamma^{\text{EM}}_{\text{A}}$, 
$\Gamma^{\text{EM}}_{\text{B}}$, and $\Gamma^{\text{EM}}_{\text{c}}$ (see also Appendix B).
Therefore, it is expected that the dynamical components of the gravitational field, i.e., graviton, will dominate the quantities
$\Gamma_{\text{A}}$, 
$\Gamma_{\text{B}}$, and $\Gamma_{\text{c}}$.
The two-point function consisting only of dynamical degrees of freedom of the gravitational field in the interaction picture is
$\langle \bigl\{\hat{h}^\text{I}_{ij} (x), \hat{h}^\text{I}_{kl} (y)\bigr\}\rangle$ with the indices $(i,j,k,l)$ running from 1 to 3.
The quantized gravitational field $\hat{h}^{\text{I}}_{ij}(x)$ is expanded into plane waves around Minkowski spacetime as follows~\cite{Suzuki,Kanno}:
\begin{align}
\hat{h}^{\text{I}}_{ij}(x)
=
\sqrt{32\pi G}
\sum_{\lambda}
\int \frac{d^3k}{\sqrt{(2\pi)^3 2k^{0}}}
\Big(
\epsilon^{\lambda}_{ij}(\bm{k})
a_{\lambda}(\bm{k})
e^{-ik_{\mu}x^{\mu}}
+
\text{H.c.}
\Big),
\end{align}
where we introduced the dispersion relation 
$k^{0}=|\bm{k}|$
and the transverse-traceless polarization tensor $\epsilon^{\lambda}_{ij}(\bm{k})$.
The index $\lambda$ represents the polarization mode of the gravitational field.
The operators $a_{\lambda}(\bm{k})$ and $a^{\dagger}_{\lambda}(\bm{k})$ are the annihilation and creation operators and obey the commutation relations
$[a_{\lambda}(\bm{k}), a^{\dagger}_{\lambda^{\prime}}(\bm{k}^{\prime})]
=\delta_{\lambda\lambda^{\prime}}\delta^{(3)}(\bm{k}-\bm{k}^{\prime})$ 
and 
$[a_{\lambda}(\bm{k}), a_{\lambda^{\prime}}(\bm{k}^{\prime})]
=
[a^{\dagger}_{\lambda}(\bm{k}), a^{\dagger}_{\lambda^{\prime}}(\bm{k}^{\prime})]=0$.
Then the two-point function 
$\langle \bigl\{\hat{h}^\text{I}_{ij} (x), \hat{h}^\text{I}_{kl} (y)\bigr\}\rangle$ 
becomes
\begin{align}
\langle \bigl\{\hat{h}^\text{I}_{ij} (x), \hat{h}^\text{I}_{kl} (y)\bigr\}\rangle
&=
32\pi G
\sum_{\lambda}
\int \frac{d^3k}{(2\pi)^3 2k^{0}}
\left[
\epsilon^{\lambda}_{ij}(\bm{k})
\epsilon^{\lambda *}_{kl}(\bm{k})
e^{-ik\cdot (x-y)}
+\text{c.c.}
\right]
\nonumber\\
\quad
&=
32\pi G
\int \frac{d^3k}{(2\pi)^3 2k^{0}}
\left[
\Pi_{ijkl}(\bm{k})
e^{-ik\cdot (x-y)}
+\text{c.c.}
\right],
\end{align}
where 
$\Pi_{ijkl}(\bm{k})$
is the sum over the polarization tensors and satisfies
$\Pi_{ijkl}(\bm{k}):=
\sum_{\lambda}
\epsilon^{\lambda}_{ij}(\bm{k})
\epsilon^{\lambda *}_{kl}(\bm{k})
=
(P_{ik}(\bm{k})P_{jl}(\bm{k})+P_{il}(\bm{k})P_{jk}(\bm{k})-P_{ij}(\bm{k})P_{kl}(\bm{k}))/2$
with the projection tensor
$P_{ij}(\bm{k})=\eta_{ij}-k_{i}k_{j}/|\bm{k}|^2$.
Using the identities
$k_{i}k_{j}/|\bm{k}|^2=\eta_{ij}/3$
and
$k_{i}k_{j}k_{k}k_{l}/|\bm{k}|^4
=
(\eta_{ik}\eta_{jl}+\eta_{il}\eta_{jk}+\eta_{ij}\eta_{kl})/15$,
we have
\begin{align}
\langle \bigl\{\hat{h}^\text{I}_{ij} (x), \hat{h}^\text{I}_{kl} (y)\bigr\}\rangle
&=
16\pi G
\int \frac{d^3k}{(2\pi)^3 2k^{0}}
\Big[
\big(
P_{ik}(\bm{k})P_{jl}(\bm{k})+P_{il}(\bm{k})P_{jk}(\bm{k})-P_{ij}(\bm{k})P_{kl}(\bm{k})
\big)
e^{-ik\cdot (x-y)}
+\text{c.c.}
\Big]
\nonumber\\
\quad
&=
\frac{32\pi G}{5}
\left(
\eta_{ik}\eta_{jl}+\eta_{il}\eta_{jk}
-\frac{2}{3}\eta_{ij}\eta_{kl}
\right)
\left[
\int \frac{d^3k}{(2\pi)^3 2k^{0}}
e^{-ik\cdot (x-y)}
+\text{c.c.}
\right]
\nonumber\\
\quad
&=
\frac{8G}{5\pi}
\left(
\eta_{ik}\eta_{jl}+\eta_{il}\eta_{jk}
-\frac{2}{3}\eta_{ij}\eta_{kl}
\right)
\lim_{\epsilon\to0}
\left[
\frac{1}{-(x^{0}-y^{0}-i\epsilon)^2+|\bm{x}-\bm{y}|^2}
+
\text{c.c.}
\right]
,
\label{twog}
\end{align}
where we used the integral
\begin{align}
\int \frac{d^3k}{(2\pi)^3 2k^{0}}
e^{-ik\cdot (x-y)}
=
\frac{1}{4\pi^2}
\lim_{\epsilon\to0}
\left[
\frac{1}{-(x^{0}-y^{0}-i\epsilon)^2+|\bm{x}-\bm{y}|^2}
\right]
\end{align}
including the UV cutoff parameter $\epsilon$.
 
\section{\label{proofEoF}PROOFS OF INEQUALITY~\eqref{EoF} AND EQ.~\eqref{discordin}}
The goal of this appendix is to prove inequality~\eqref{EoF} and Eq.~\eqref{discordin}.
To achieve this, it is convenient to introduce the Koashi-Winter relation~\cite{Koashi} of a pure tripartite system 
$|\Psi_{\text{ABE}}\rangle$ 
as follows:
\begin{align}
S(\rho_{\text{A}})
=
E_{\text{f}}(\rho_{\text{AE}})+\mathcal{J}(\text{A}, \text{B}),
\label{KW}
\end{align}
where the entanglement of formation 
$E_{\text{f}}(\rho_{\text{AE}})$ 
is defined by
\begin{align}
E_{\text{f}}(\rho_{\text{AE}})
:=
\min_{\{p_{i},|\psi_{\text{AE}}\rangle_{i}\}}
\sum_{i}p_{i}S(\Tr_{\text{E}}[|\psi_{\text{AE}}\rangle_{i}\langle\psi_{\text{AE}}|])
\end{align}
with states $|\psi_{\text{AE}}\rangle_{i}$ due to Schmidt decomposition
$|\Psi_{\text{ABE}}\rangle
=\sum_{i}\sqrt{p_{i}}|\psi_{\text{AE}}\rangle_{i}\otimes|\psi_{\text{B}}\rangle_{i}$ 
satisfying $\sum_{i}p_{i}=1$ and $p_{i}\geq0$.
The minimization is taken over all ensembles 
$\{p_{i},|\psi_{\text{AE}}\rangle_{i}\}$ 
such that $\sum_{i}p_{i}|\psi_{\text{AE}}\rangle_{i}\langle\psi_{\text{AE}}|
=\rho_{\text{AE}}$.
Roughly speaking, this entanglement of formation characterizes at least how many maximally entangled states $|\psi_{\text{AE}}\rangle$ required to generate the state 
$\Tr_{\text{E}}[|\psi_{\text{AE}}\rangle\langle\psi_{\text{AE}}|]$. 
$\mathcal{J}(A, B)$ in the second term of Eq.~\eqref{KW} is the classical correlation, which is seen as the amount of information about subsystem A that can be obtained by performing a measurement on subsystem B, and is defined by
\begin{align}
\mathcal{J}(\text{A}, \text{B})
:=
S(\rho_{\text{A}})-
\min_{\{\Pi_{i}\}}\sum_{i}p_{i}S(\rho_{\text{A}|i}),
\end{align}
where $S(\rho_{\text{A}|i})$ is the von Neumann entropy of the post-measurement state $\rho_{\text{A}|i}$ with the probability ${p_{i}}$ defined as
\begin{align}
\rho_{\text{A}|i}
:=
\frac{1}{p_{i}}\text{Tr}_{\text{B}}\left[(\mathbb{1}_{\text{A}}
\otimes
\Pi^{\text{B}}_{i})\rho_{\text{AB}}(\mathbb{1}_{{\text{A}}}\otimes\Pi^{\text{B}}_{i})\right]
,
\quad
p_{i}
:=
\text{Tr}_{\text{AB}}
\left[
(\mathbb{1}_{\text{A}}\otimes\Pi^{\text{B}}_{i})
\rho_{\text{AB}}
(\mathbb{1}_{\text{A}}\otimes\Pi^{\text{B}}_{i})
\right].
\end{align}
$\Pi_{i}$ is the positive operator valued measure (POVM) acting on the subsystem B.
The condition $\min_{\{\Pi_{i}\}}$ is introduced not to disturb all states, i.e., we must choose the projective operator $\{\Pi_{i}\}$ so as to reduce the dependence on the projection measure. 
Note that in contrast to classical theory, the measurement in subsystem B disturbs the subsystem A. 
When we measure the state of subsystem B, the wave function collapses, and the state of subsystem B is determined; that is, the projective measure makes a condition to the state of subsystem A.

The classical correlation $\mathcal{J}(\text{A}, \text{B})$ is related to the quantum discord $\mathcal{D}(\text{A}, \text{B})$\cite{Ollivier}.
The definition of quantum discord is the difference between the quantum mutual information 
$\mathcal{I}(\text{A}, \text{B})
=
S(\rho_{\text{A}})+S(\rho_{\text{B}})-S(\rho_{\text{AB}})$
and the classical correlation 
$\mathcal{J}(\text{A}, \text{B})$:
\begin{align}
\mathcal{D}(\text{A}, \text{B})
=
\mathcal{I}(\text{A}, \text{B})-\mathcal{J}(\text{A}, \text{B}).
\label{discord}
\end{align}
The quantum mutual information 
$\mathcal{I}(\text{A}, \text{B})$ 
quantifies the total amount of correlations between the two subsystems A and B.
We note that the quantum mutual information is always non-negative due to the subadditivity of von Neumann entropy.
In classical theory, $\mathcal{D}(\text{A}, \text{B})=0$ is always correct; however, in quantum theory, it can become $\mathcal{D}(A, B)>0$.

By using Eq.~\eqref{KW} and~\eqref{discord}, we can prove the inequality 
$E_{\text{f}}(\rho_{\text{AE}})\geq S(\text{A}|\text{B})$
as follows:
\begin{align}
E_{\text{f}}(\rho_{\text{AE}})
&=
S(\rho_{\text{A}})-\mathcal{J}(A, B)
\nonumber\\
\quad
&=
S(\rho_{\text{A}})-\mathcal{I}(\text{A}, \text{B})+\mathcal{I}(\text{A}, \text{B})-\mathcal{J}(\text{A}, \text{B})
\nonumber\\
\quad
&=
S(\rho_{\text{A}})-\mathcal{I}(\text{A}, \text{B})+\mathcal{D}(\text{A}, \text{B})
\nonumber\\
\quad
&
\geq
S(\rho_{\text{A}})-\mathcal{I}(\text{A}, \text{B})
\nonumber\\
\quad
&=
S(\rho_{\text{AB}})-S(\rho_{\text{B}})
=
S(\text{A}|\text{B}),
\label{AEB}
\end{align}
where we inserted the quantum mutual information 
$\mathcal{I}(\text{A}, \text{B})$ 
in the second line and 
$\mathcal{D}(\text{A}, \text{B})\geq0$ 
was used in the fourth line.
Another reordered version of the inequality~\eqref{AEB} is computed as
\begin{align}
E_{\text{f}}(\rho_{\text{AB}})
&\geq
S(\rho_{\text{AE}})-S(\rho_{\text{E}})
\nonumber\\
\quad
&=
S(\rho_{\text{B}})-S(\rho_{\text{AB}})
\nonumber\\
\quad
&=
-S(\text{A}|\text{B}),
\label{ABE}
\end{align}
where, in the second line, we used the properties 
$S(\rho_{\text{E}})=S(\rho_{\text{AB}})$ 
and 
$S(\rho_{\text{AE}})=S(\rho_{\text{B}})$, which holds because the state $|\psi_{\text{ABE}}\rangle$ is pure state.
Note that these properties are always satisfied because of the invariance of the von Neumann entropy under unitary evolution when the initial state is in a pure state.
In the last line, we inserted the definition of the conditional von Neumann entropy $S(\text{A}|\text{B}):=S(\rho_{\text{AB}})-S(\rho_{\text{B}})$.

Furthermore, we can show the equation $\mathcal{D}(\text{A,E})=E_{\text{f}}(\rho_{\text{AB}})+S(\text{A}|\text{B})$ as follows:
\begin{align}
\mathcal{D}(\text{A,E})
=
E_{\text{f}}(\rho_{\text{AB}})+S(\rho_{\text{E}})-S(\rho_{\text{AE}})
=
E_{\text{f}}(\rho_{\text{AB}})+S(\rho_{\text{AB}})-S(\rho_{\text{B}})
=
E_{\text{f}}(\rho_{\text{AB}})+S(\text{A}|\text{B}),
\end{align}
where, in the first equality, we used another reordered version of the Koashi-Winter relation~\eqref{KW} with respect to B and E:
\begin{align}
S(\rho_{\text{A}})
&=
E_{\text{f}}(\rho_{\text{AB}})+\mathcal{J}(\text{A,E})
\nonumber\\
\quad
&=
E_{\text{f}}(\rho_{\text{AB}})+\mathcal{I}(\text{A,E})-\mathcal{I}(\text{A,E})+\mathcal{J}(\text{A,E})
\nonumber\\
\quad
&=
E_{\text{f}}(\rho_{\text{AB}})+\mathcal{I}(\text{A,E})-\mathcal{D}(\text{A,E}).
\end{align}
In the inequality Eq.~\eqref{EoF} and Eq.~\eqref{discordin}, state E is regarded as the graviton state.
Therefore, inequality Eq.~\eqref{EoF} and Eq.~\eqref{discordin} has been proven.

\end{appendix}

\end{document}